\documentclass[a4paper,fleqn,usenatbib]{mnras}
\usepackage{mathptmx}
\usepackage[T1]{fontenc}
\usepackage{ae,aecompl}
\usepackage{graphicx}	
\usepackage{amsmath}	
\usepackage{amssymb}	
\usepackage{threeparttable}
\usepackage{color}
\usepackage{bm}
\usepackage{epsfig}
\usepackage{wrapfig}
\usepackage{multicol}
\usepackage{exscale}
\usepackage{enumitem}

\title[Star Formation in Luminous Quasars at $2<z<3$]{Star Formation Rates in Luminous Quasars at $2<z<3$}

\author[K.A.Harris et al.]{Kathryn Harris$^{1,2,3}$\thanks{E-mail: kateharris142@gmail.com}, 
Duncan Farrah$^{1}$, 
Bernhard Schulz$^{4,5}$, 
Evanthia Hatziminaoglou$^{6}$, \newauthor 
Marco Viero$^{7}$,
Nick Anderson$^{1}$,
Matthieu B{\'e}thermin$^{6}$,
Scott Chapman$^{8}$, \newauthor 
David L. Clements$^{9}$,
Asantha Cooray$^{10}$,
Andreas Efstathiou$^{11}$,
Anne Feltre$^{12}$, \newauthor 
Peter Hurley$^{13}$, 
Eduardo Ibar$^{14}$,
Mark Lacy$^{15}$, 
Sebastian Oliver$^{13}$, 
Mathew J. Page$^{16}$,\newauthor 
Ismael P\'erez-Fournon$^{2,3}$, 
Sara M. Petty$^{1}$,
Lura K. Pitchford$^{1,6}$,  
Dimitra Rigopoulou$^{17}$, \newauthor
Douglas Scott$^{18}$,
Myrto Symeonidis$^{16}$,  
Joaquin Vieira$^{19}$,  
Lingyu Wang$^{20,21}$ \\
$^{1}$ Department of Physics, Virginia Tech, Blacksburg, VA 24061, USA \\
$^{2}$ Instituto de Astrof{\'i}sica de Canarias, C/via L{\'a}ctea s/n, E-38205 La Laguna, Tenerife, Spain \\ 
$^{3}$ Universidad de La Laguna, Dpto. Astrof{\'i}sica, E-38206 La Laguna, Tenerife, Spain \\
$^{4}$ California Institute of Technology, 1200 E. California Blvd., Pasadena, CA 91125, USA \\ 
$^{5}$ Infrared Processing and Analysis Center, MS 100-22, California Institute of Technology, JPL, Pasadena, CA 91125, USA \\
$^{6}$ European Southern Observatory, Karl-Schwarzschild-Str. 2, 85748 Garching, Germany \\
$^{7}$ Kavli Institute for Particle Astrophysics and Cosmology, Stanford University, 382 Via Pueblo Mall, Stanford, CA 94305, USA \\
$^{8}$ Department of Physics and Atmospheric Science, Dalhousie University, Halifax, NS B3H 3J5, Canada \\
$^{9}$ Astrophysics Group, Imperial College London, Blackett Laboratory, Prince Consort Road, London SW7 2AZ, UK \\
$^{10}$ Department of Physics \& Astronomy, University of California, Irvine, CA 92697, USA\\
$^{11}$ School of Sciences, European University Cyprus, Diogenes Street, Engomi, 1516 Nicosia, Cyprus\\
$^{12}$ Sorbonne Universit\'es, UPMC-CNRS, UMR7095, Institut d'Astrophysique de Paris, F-75014, Paris, France\\
$^{13}$ Astronomy Centre, Department of Physics and Astronomy, University of Sussex, Falmer, Brighton BN1 9QH, UK \\
$^{14}$ Instituto de Fisica y Astronomia, Universidad de Valparaiso, Avda. Gran Bretana 1111, Valparaiso, Chile \\
$^{15}$ National Radio Astronomy Observatory, 520 Edgemont Road, Charlottesville, VA 22903, USA \\
$^{16}$ Mullard Space Science Laboratory, University College London, Holmbury St. Mary, Dorking, Surrey RH5 6NT, UK\\
$^{17}$ Department of Physics, University of Oxford, Keble Road, Oxford OX1 3RH, UK\\
$^{18}$ Physics \& Astronomy, University of British Columbia, 6224 Agricultural Road, Vancouver, BC V6T 1Z1 \\
$^{19}$ Department of Astronomy and Department of Physics, University of Illinois, 1002 West Green St., Urbana, IL 61801\\
$^{20}$ SRON Netherlands Institute for Space Research, Landleven 12, 9747 AD, Groningen, The Netherlands\\
$^{21}$ Institute for Computational Cosmology, Department of Physics, Durham University, Durham, DH1 3LE, UK
}

\date{Accepted February 2nd 2016.}

\pubyear{2016}

\begin{document}
\label{firstpage}
\pagerange{\pageref{firstpage}--\pageref{lastpage}}
\maketitle

\begin{abstract}
We investigate the relation between star formation rates ($\dot{\rm{M}}_s$) and AGN properties in optically selected type 1 quasars at $2<z<3$ using data from \textit{Herschel} and the SDSS. We find that $\dot{\rm{M}}_s$ remains approximately constant with redshift, at $300\pm100~\rm{M}_{\odot}$yr$^{-1}$. Conversely, $\dot{\rm{M}}_s$ increases with AGN luminosity, up to a maximum of $\sim600~\rm{M}_{\odot}$yr$^{-1}$, and with \ion{C}{IV} FWHM. In context with previous results, this is consistent with a relation between $\dot{\rm{M}}_s$ and black hole accretion rate ($\dot{\rm{M}}_{bh}$) existing in only parts of the $z-\dot{\rm{M}}_{s}-\dot{\rm{M}}_{bh}$ plane, dependent on the free gas fraction, the trigger for activity, and the processes that may quench star formation. The relations between $\dot{\rm{M}}_s$ and both AGN luminosity and \ion{C}{IV} FWHM are consistent with star formation rates in quasars scaling with black hole mass, though we cannot rule out a separate relation with black hole accretion rate. Star formation rates are observed to decline with increasing \ion{C}{iv} equivalent width. This decline can be partially explained via the Baldwin effect, but may have an additional contribution from one or more of three factors; $M_i$ is not a linear tracer of L$_{2500}$, the Baldwin effect changes form at high AGN luminosities, and high \ion{C}{iv} EW values signpost a change in the relation between $\dot{\rm{M}}_s$ and $\dot{\rm{M}}_{bh}$. Finally, there is no strong relation between $\dot{\rm{M}}_s$ and Eddington ratio, or the asymmetry of the \ion{C}{iv} line. The former suggests that star formation rates do not scale with how efficiently the black hole is accreting, while the latter is consistent with \ion{C}{iv} asymmetries arising from orientation effects.
\end{abstract}

\begin{keywords}
galaxies: evolution, starburst -- quasars: general -- infrared: galaxies
\end{keywords}

\section{Introduction}\label{sec:intro}
Three lines of evidence suggest that, at all redshifts, there is a deep connection between star formation rates in galaxies, and the presence and properties of Active Galactic Nuclei (AGN) in their centres. First, there is a strong similarity in the evolution of comoving luminosity density of AGN and star formation \citep[e.g.][]{Madau2014}. For AGN, the optical luminosity function of quasars plateaus between $2 < z < 3$ \citep[e.g.][]{Richards2006,Delvecchio2014}. For star formation, the comoving star formation rate density increases by a factor of at least ten over $0 < z < 1$ \citep[e.g.][]{Lilly1996,Dickinson2003,Merloni2004,wangli13}, peaks at $z\sim2$ \citep[e.g.][]{Connolly1997,Somerville2001,Lanzetta2002,hop06} and then declines at higher redshifts \citep[e.g.][]{PerezGonzalez2005,wall08,Bethermin2012,Wuyts2011}. Second, there is a positive relationship between stellar and supermassive black hole (hereafter just black hole) mass in quiescent galaxies, with more massive black holes being found in systems with, on average, a higher bulge stellar mass \citep[e.g.][]{Magor98,Tremaine2002}. Finally, observations find luminous star formation and AGN in the same galaxies at rates significantly higher than expected by chance \citep[e.g.][]{gen98,Farrah2003,Alexander2005,lon06,cop10,Mainieri2011,wang13,far13,cas14,pit15}. 

The scaling relations, or lack thereof, between star formation rates and AGN properties as a function of variables such as redshift, luminosity, stellar mass and environment, are fundamental to understanding the nature of this connection, and by extension to  understanding the mass assembly history of galaxies throughout the history of the Universe. The existence and nature of such scaling relations give insights into how initially free baryons are converted into stars and black holes, and how these conversions are affected by factors such as free gas availability. They also give insights into more apparent manifestations of this connection, such as the idea that star formation and AGN activity can directly affect each other. One example of this is `AGN quenching', where winds or jets from an AGN curtail star formation in its host galaxy on timescales much shorter than gas depletion timescales (see e.g. \citealt{Fabian2012} for a review). AGN quenching is motivated both by models \citep{croton06,somer08,hir14,sch15}, which in many cases require quenching to bring their predictions in line with observations, and by some observations that find massive outflows of gas that are plausibly linked to both AGN winds, and reductions in star formation rates \citep{chung11,Farrah2012,Trichas2012,gui15}. Another example is the suggestion that, in some circumstances, the AGN can trigger star formation \citep[e.g.][]{king05,ish12,gaib12,silk13,Zubovas2013}. 

In this paper we explore the connection between AGN activity and star formation in optically luminous type 1 quasars at $2<z<3$. We choose this class of object since they are straightforward to find, correspond to a specific stage in the AGN duty cycle, and reside at the epoch where the comoving luminosity density arising from both star formation and AGN activity is expected to peak. We assemble our sample from the Sloan Digital Sky Survey, restricted to fields with high quality far-infrared imaging, from which we derive star formation rates. We compare host star formation rates to the properties of the AGN as derived from the SDSS data to see how star formation rates depend on AGN properties across the quasar population. We adopt AB magnitudes, and assume a spatially flat cosmology with $\Omega_{\rm{m}} = 0.3$ and $H_{0}=70\,$km s$^{-1}$ Mpc$^{-1}$.

\begin{figure} 
\includegraphics[width=0.47\textwidth]{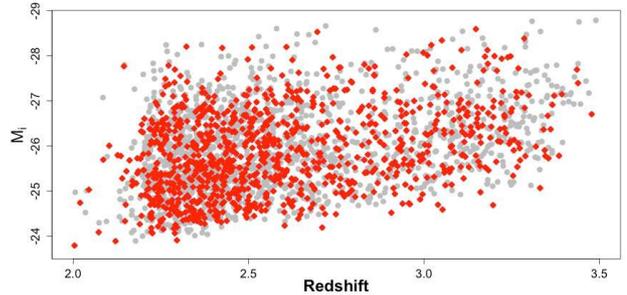}
\caption{The $z - M_i$ plane showing both all SDSS DR9 quasars in Stripe 82 with uniform flag $>0$ and $2.15<z<3.5$ (grey points), and our sample of quasars (red points). The distributions of the red and grey points are identical by visual inspection, and indistinguishable via a K-S test (\S\ref{ssectsample}).}
\label{fig:zImagComp} 
\end{figure}

\section{Data}\label{sectdata}

\subsection{Sample Selection}\label{ssectsample}
We selected our sample from the Baryonic Oscillation Spectroscopic Survey (BOSS), itself part of the Sloan Digital Sky Survey Data Release 9 Quasar catalogue (SDSS DR9Q, \citealt{Paris2012}). By selecting from BOSS, we can assemble the necessary number of quasars to subdivide into multiple bins based on various criteria, and still have sufficient objects per bin to perform stacking analyses (see \S\ref{bossanalysis}). 

We start with the 61,931 SDSS DR9 quasars at $z>2.15$. Our selection comprises four steps. First, we excluded all BOSS quasars that did not lie within Stripe 82. This left 9,520 objects. Second, we excluded all quasars with a uniform flag of zero or below, leaving only those with a uniform flag of 1 or 2. The remaining quasars are the CORE quasar selection within BOSS. The CORE selection is both homogeneous and uniform, and is highly, but not entirely complete over the range $2.15 \leq z \leq 3.5$ \citep{Ross2012,White2012}. We discuss the issue of completeness further in \S\ref{caveatsoi}. This step left 6,516 objects. Third, we excluded all quasars at $z>3.5$ and $z<2.15$, leaving 3,256 objects. Fourth, we excluded quasars that lie outside or within 20 pixels of the edge of the HerS and/or HeLMS data (\S\ref{ssectfirdata}), so that there were sufficient pixels in the stacked images to allow for clustering corrections (\S\ref{ssectstack}). The pixel sizes differ for each band, with the largest being for $500\, \mu$m. We therefore used the $500\, \mu$m image for this step, which left 1,002 objects. Our selection does not include steps to remove either Broad Absorption Line (BAL) quasars, or sources with radio data. The effects of this decision are described in \S\ref{caveatsiii}. 

\begin{figure}
\begin{tabular}{c}
  \vspace{-0.3cm}
  \includegraphics[width=0.40\textwidth]{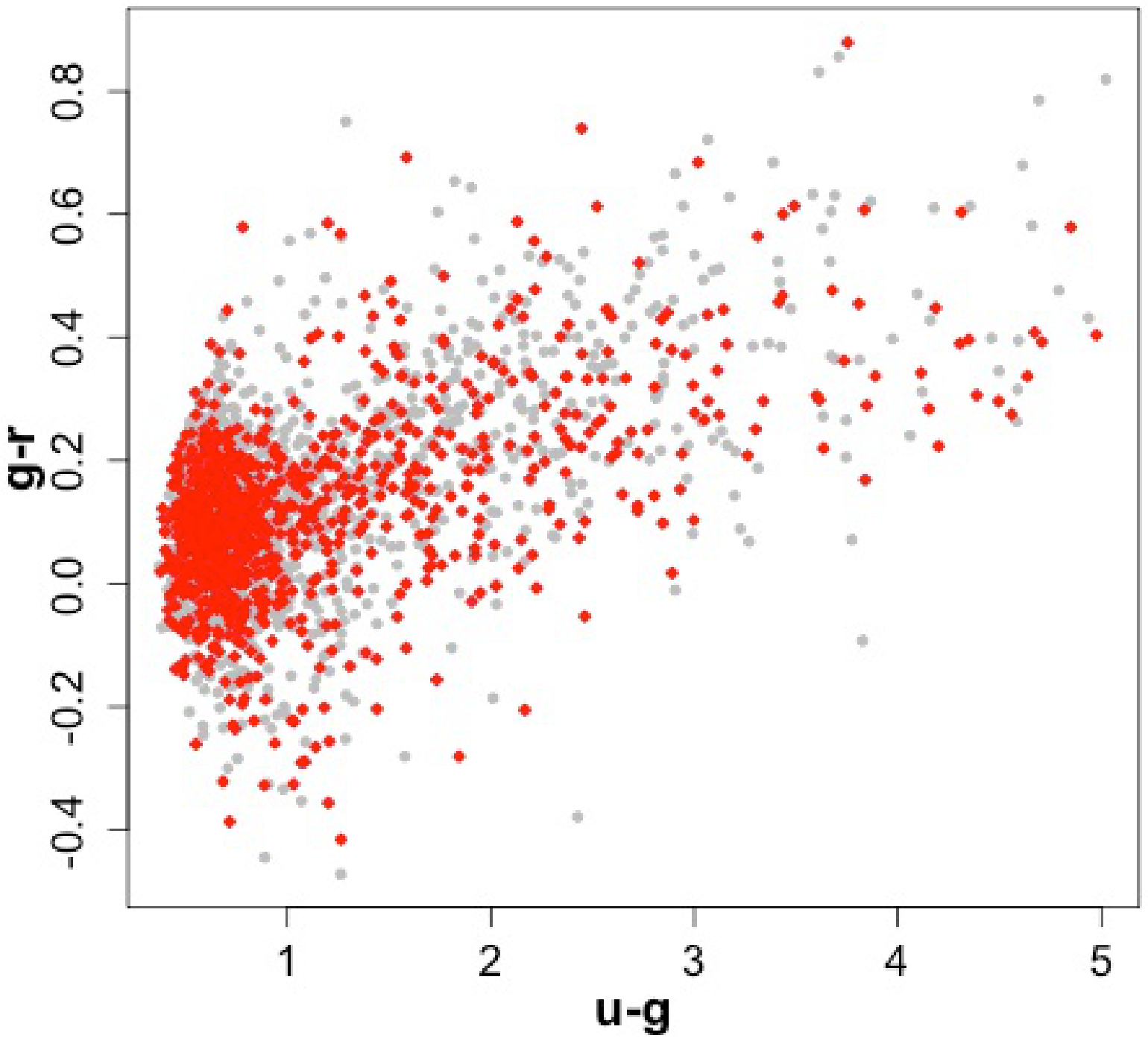} \\
  \vspace{-0.3cm}
  \includegraphics[width=0.40\textwidth]{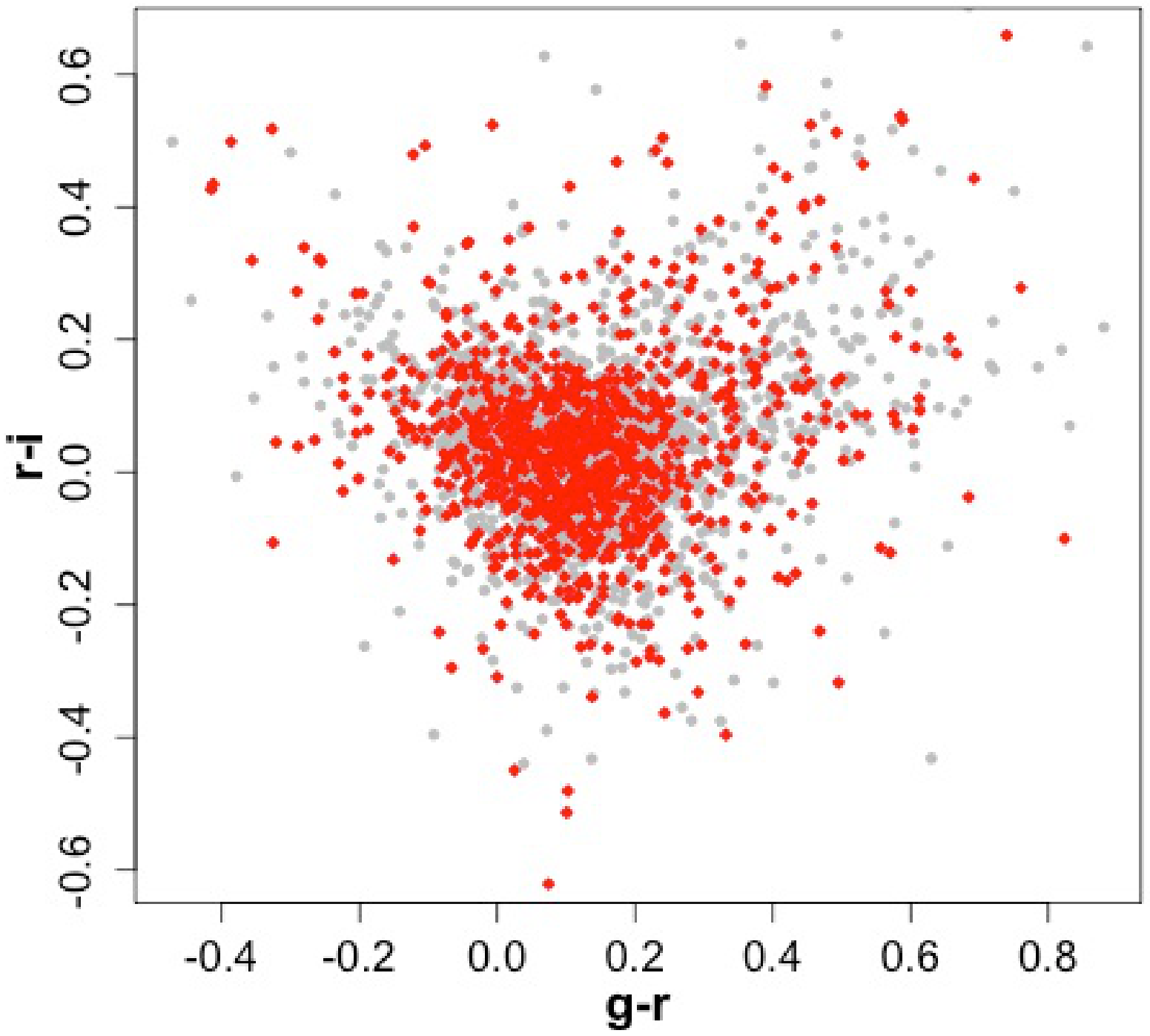} \\
  \vspace{-0.3cm}
  \includegraphics[width=0.40\textwidth]{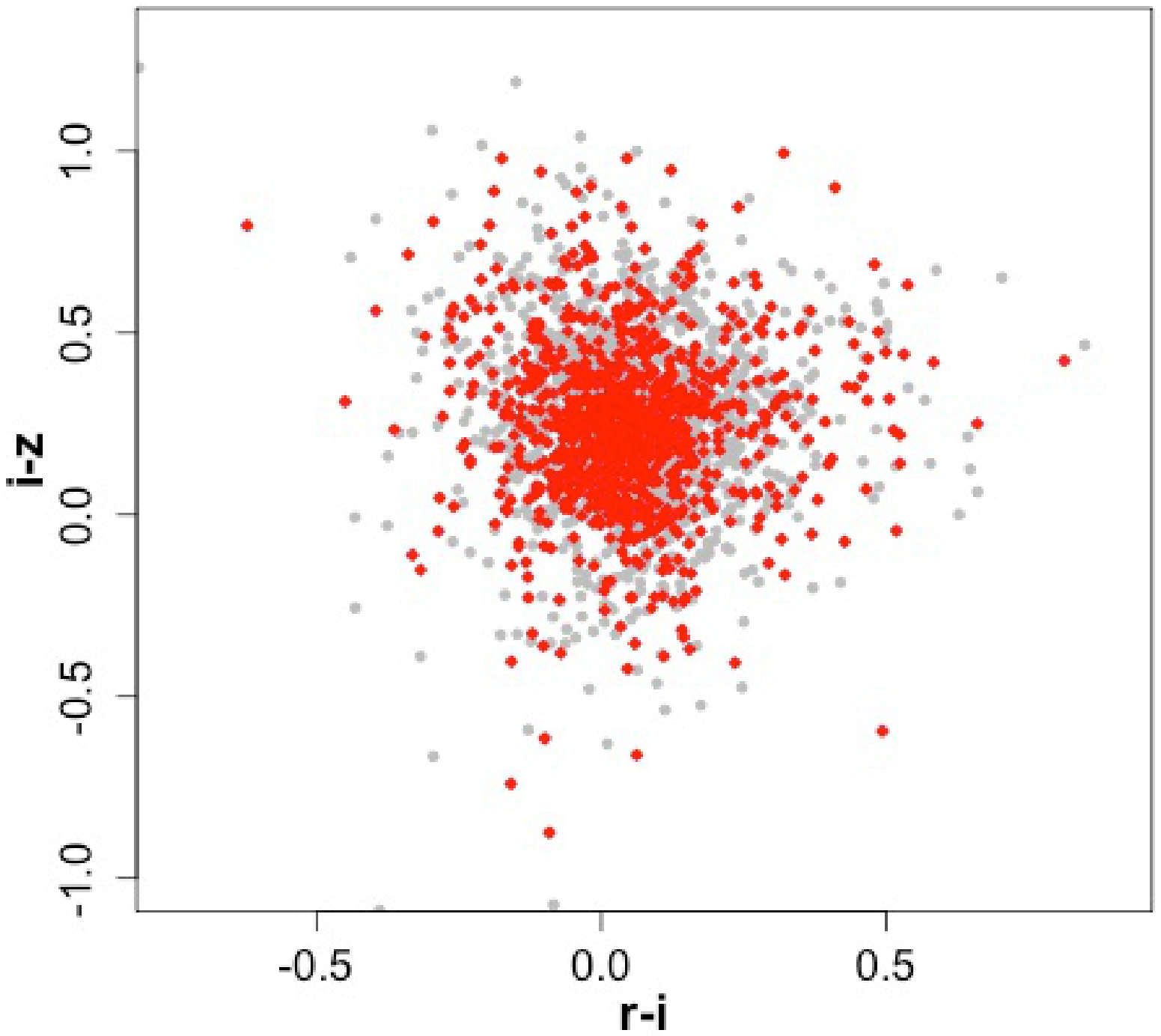} 
\end{tabular}
\caption{Observed-frame colour-colour diagrams for our sample (red) and all SDSS DR9 quasars in Stripe 82 with uniform flag $>0$ and $2.15<z<3.5$ (grey). The distributions are identical by visual inspection, and indistinguishable via a K-S test.}
\label{fig:colorComp} 
\end{figure}

\begin{figure} 
\includegraphics[width=0.50\textwidth]{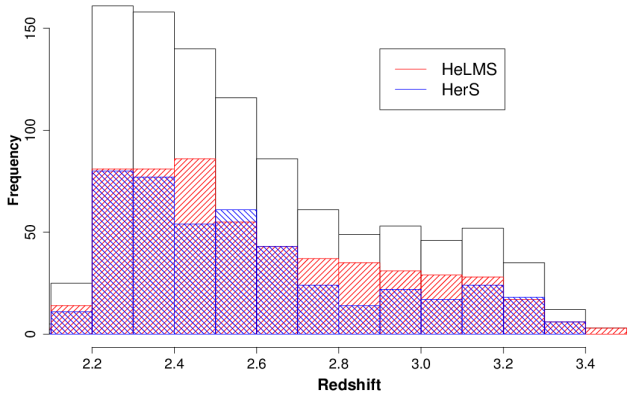}
\caption{Redshift distributions for the quasars in HeLMS and HerS (red and blue hatched histograms, respectively) and for all SDSS DR9 quasars in Stripe 82 with uniform flag $>0$ and $2.15<z<3.5$ (white solid).}
\label{fig:zhist} 
\end{figure}

We checked how our sample compared to all of the SDSS BOSS quasars within Stripe 82. The absolute $i$ band magnitudes as functions of redshift of the two samples are qualitatively identical (Fig. \ref{fig:zImagComp}). A KS test yields a p-value of 0.685; high enough to accept the null hypothesis that they are from the same parent population. Comparing their SDSS colours (Fig. \ref{fig:colorComp}) again gives qualitatively identical distributions, and $p$-values high enough to accept that they are drawn from the same parent population ($\geq0.18$ in all cases). Their redshift distributions (Fig. \ref{fig:zhist}) show no differences, and yield a $p$-value of 0.86. We repeated these tests with the HerS and HeLMS fields separately, and found no significant differences. We conclude that our sample is representative of the SDSS BOSS Stripe 82 quasar population at $2.15<z<3.5$.

\subsection{Far-infrared Imaging}\label{ssectfirdata}
We assembled far-infrared imaging data from the \textit{Herschel} Stripe 82 Survey (HerS; \citealt{Viero2014}) and the HerMES Large-Mode Survey (HeLMS; \citealt{Oliver2012}, map version 0.2). Both surveys were performed using the Spectral and Photometeric Imaging REceiver instrument (SPIRE; \citealt{Griffin2010}) onboard the \textit{Herschel} Space Observatory \citep{Pilbratt2010}, in the $250\, \mu$m, $350\, \mu$m and $500\, \mu$m bands. Both of the surveys form part of the \textit{Herschel} Multi-tiered Extragalactic Survey (HerMES; \citealt{Oliver2012}). HeLMS cover 270 deg$^2$ and HerS covers 79 deg$^2$, with an overlap between the two of $6$ deg$^2$. The HerS and HeLMS surveys together cover 115 of the 255 deg$^2$ SDSS Stripe 82 field \citep{abaza09}. \textit{Herschel} map-making details can be found in \citet{Patanchon2008} and \citet{Viero2014}.

\section{Analysis}\label{bossanalysis}
The high redshifts of our sample mean they are virtually all individually undetected by \textit{Herschel}. Of the 1,002 quasars, only $\sim 5\%$ have a catalogue flux density in at least one band, and in all cases those flux densities are close to the detection limit. To measure their star formation rates we thus stack the \textit{Herschel} data for sets of objects selected according to specific criteria, and then fit model spectral energy distributions (SEDs) to the stacked flux densities to extract mean star formation rates for that set. We describe the stacking and SED fitting procedures in the following subsections.

\subsection{Stacking}\label{ssectstack}
We stacked our sample as a function of six variables, all taken from \citet{Paris2012}: redshift; absolute $i$ band magnitude ($M_i$); $\Delta[g-i]$ colour; and the (rest-frame) equivalent width (EW), Full Width at Half Maximum (FWHM), and FWHM asymmetry of the \ion{C}{iv}$\lambda$1549\AA\ line (hereafter we refer to this line as \ion{C}{iv}). The $M_i$ values include a correction for Galactic extinction and assume a power law continuum index of $\alpha=-0.5$. The K-corrections are pegged to the $i$-band for a quasar at $z=2$ rather than $z=0$ as K-correcting to a redshift close to the median redshift of the sample substantially reduces systematic error. The consequence is that the $M_i$ values are distance-corrected to 10pc but are K-corrected to the equivalent of the $i$-band filter observing at $z=2$, i.e. rest-frame 2500\AA\ . The K-corrections are computed using the values in table 4 of \citet{Richards2006}. The $\Delta[g-i]$ values are the Galactic extinction corrected {\itshape difference} in observed-frame $g-i$ colour between a quasar and the mean for all SDSS DR9 quasars at the same redshift. The \ion{C}{iv} asymmetry is the ratio of the blue-side FWHM to the red-side FWHM. 

We found that each {\itshape Herschel} stack needed a minimum of 20 quasars to give detections or useful limits. We thus arranged the bin spacing of the variables to ensure at least 20 quasars per bin as a first priority, and to be approximately equally spaced as a second priority. The resulting bins were as follows:

\begin{itemize}
\item $z$: eleven bins with mean values: 2.205, 2.286, 2.370, 2.452, 2.540, 2.639, 2.746, 2.854, 2.970, 3.108, 3.266. These intervals correspond to $\Delta t = 0.1\,$Gyr. 
\item $M_i$: eight bins with mean values: -24.33, -24.75, -25.25, -25.75, -26.25, -26.75, -27.25, and -27.88. 
\item $\Delta[g-i]$: eight bins with mean values: -0.34, -0.24, -0.14, -0.05, 0.05, 0.15, 0.25, 0.35. The histogram of $\Delta[g-i]$ is approximately a gaussian centered at zero with a FWHM of $\sim0.16$. Hence, to ensure $>20$ objects per bin we restrict the $\Delta[g-i]$ to this range. We do however explore $\Delta[g-i]$ values outside this range in \S\ref{resabscol}. 
\item \ion{C}{iv} FWHM ($F_{\rm{C}}$): ten bins with mean values: 2467, 2991, 3474, 3858, 4200, 4543, 4873, 5281, 5859, and 6986 km\,s$^{-1}$.
\item \ion{C}{iv} EW ($E_{\rm{C}}$): eleven bins with mean values: 20.48, 28.59, 33.50, 37.76, 41.71, 47.52, 54.95, 63.92, 78.40, 96.16, and 136.30 nm.
\item \ion{C}{iv} FWHM asymmetry ($A_{\rm{C}}$): ten bins with mean values: 0.492, 0.607, 0.679, 0.740, 0.804, 0.862, 0.915, 0.987, 1.108, and 1.468.
\end{itemize} 

\noindent To perform the stacking we followed the approach of previous authors \citep{Bethermin2010,Bethermin2012,Heinis2013}. Sub-images of $41\times41$ pixels were extracted around each object and placed in a data-cube. This sub-image size gives a spatial scale, at $z=2$, of  $\sim$2 Mpc for the 250$\mu$m band and $\sim$4 Mpc for the 500$\mu$m band. The stacked flux density in each pixel is then the mean of that column in the data-cube. The flux density is found by fitting to the stacked profile a model point-spread function (PSF) for SPIRE, as implemented within the \textit{Herschel} Interactive Processing Environment (HIPE) v12 \citep{ott10}. The random error for each stacked pixel was calculated via bootstrap resampling; multiple samples of images were selected without withdrawal from each data-cube, stacked, and the resulting pixel flux densities calculated. In the six deg$^2$ overlap region between HerS and HeLMS, we used only the HeLMS data, which are deeper, since adding the HerS data to the HeLMS data made a negligible difference to the results.

\subsection{Clustering Correction}\label{ssecterrbudg}
The coarse spatial resolution of SPIRE means that flux density measurements of individual sources may be boosted due to the presence of other, individually undetected sources in the area covered by the SPIRE beam. It is however straightforward to derive a reasonably accurate correction for this effect. Star formation in high redshift systems typically spans scales of $\lesssim30$kpc \citep[e.g.][]{carn13,wiklind14,simp15}, or $\lesssim3.72$\arcsec\, at $z=2.5$. This star formation will be unresolved by SPIRE. Thus, if host galaxy star formation alone is present in the \textit{Herschel} beam, the detections in the stacked images will be point sources. Any deviations from a point source profile can thus be ascribed to contributions from other far-infrared-emitting sources around the quasars.  

To correct for these sources we compared the profiles of the stacked images to that of the  SPIRE PSF. We found that in all cases there was a small but clear excess over a pure PSF profile. To model these excesses we fitted a model consisting of the SPIRE PSF plus a  power law with slope of 1.8 to each stacked profile \citep{Bethermin2010,Bethermin2012,Heinis2013}. The power law profiles are then the clustering correction for each stack, and were removed. The sizes of the clustering corrections ranged from $\sim10\%$ for 250$\mu$m to $\sim22\%$ for 500$\mu$m.

We checked the effect of image cutout size on the derived clustering correction by repeating the clustering analysis, starting with a size of $11\times11$ pixels and increasing it until no effect on the clustering correction was seen. We found that $41\,\times\,41$ pixels was the smallest size that produced a clustering correction not dependent on the image size, and so adopted this size for the stacking.

\subsection{Star Formation Rates}\label{measrates}
We estimate star formation rates by fitting the SPIRE data with radiative transfer models for a star-forming region \citep{Efstathiou2000,Efstathiou2009}. There is substantial theoretical \citep[e.g.][]{Pier1992,Fritz2006,Schartmann2008} and observational \citep[e.g.][]{Schweitzer2006,Netzer2007,hatz10,Shao2010,Mullaney2011,Ros12,Magdis2013,Delvecchio2014,zak16} evidence that radio-quiet AGN are at least $0.5-1$ dex less luminous than star formation at $\lambda_{rest}>60\mu$m. Since the SPIRE bands always sample $\lambda_{rest}>80\mu$m at $z<3$, it is plausible that the contribution from AGN-heated dust to the SPIRE flux densities is insignificant. We do however explore the possibility of an AGN contribution in \S\ref{caveatsii}.

\begin{figure}
\includegraphics[width=0.35\textwidth,angle=90]{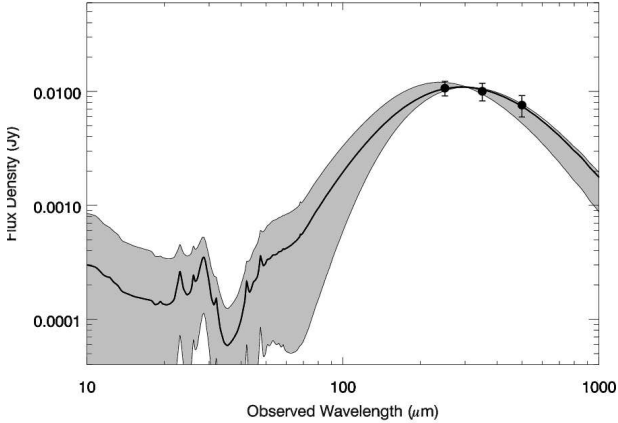} 
\caption{An example fit of a starburst model to the SPIRE fluxes extracted from the stacked data (\S\ref{measrates}). This fit is for the seventh $M_i$ bin (see supplementary online materials). The black line shows the best-fit SED, while the shaded region shows the range in SED shapes that are consistent with the range in starburst luminosity for this fit.} 
\label{fig:expuresb} 
\end{figure}

The fits are excellent, with $\chi^2_{red}<1$ in all cases. An example fit is shown in Fig. \ref{fig:expuresb}. We determine starburst luminosities and their errors by combining the acceptable fits into a weighted probability distribution function (PDF) and then extracting the peak and 68\% confidence interval of the PDF. As such, the errors include uncertainty arising from both the formal error on the best fit, as well as from the range in acceptable SED shapes and the SPIRE absolute calibration error. In all cases the PDFs are consistent with gaussian profiles. 

The starburst models vary in age, initial optical depth of the molecular clouds, and e-folding time, $\tau$, of the star formation rate (assuming a decline of the form $e^{-t/\tau}$) over the ranges 0-70~Myr, 50-125, and 10-40~Myr, respectively. With only three flux densities we cannot constrain these parameters. We adopt this approach rather than fitting a modified blackbody for two reasons. First, using a reasonable range of starburst SED shapes gives a better estimate of the uncertainties on the star formation rates. Second, a modified blackbody does not include mid-IR emission from PAHs and hot dust species and hence underestimates the IR emission due to star formation by approximately 10\%.

\subsection{Sources of Uncertainty}\label{ssecterrbudgb}
We considered five potential sources of uncertainty to assess their impact on subsequent analyses. First, we checked to see if a small number of objects were unduly affecting the stacked flux densities. To do so we tested each stack with jackknife resampling. The jackknife test was run 20 times on a random selection of samples. Each stack was split into two equal size subsamples, stacked, and the flux density measured. The values for each half were in all cases within the errors of the value for the full stack. Second, we inspected the optical spectra of all our sample for signs of contaminating sources along their lines of sight, but found none. Third, we assumed that no colour corrections were required, since such colour corrections are negligible for objects with a quasar spectral index \citep{Griffin2010}. Fourth, we did not attempt to correct for contributions from Galactic dust emission \citep[e.g.][]{Wang15}. If Galactic dust does contribute, it would systematically overestimate the flux densities by at most 10\%  at 250$\mu$m, and 30\% at 500$\mu$m. Fifth, we have no {\itshape a priori} knowledge of the actual distribution of the \textit{Herschel} flux densities of our sample. Thus, we do not know how close the mean signal from the stacks is to the mode, or typical signal, of the population. The results from the jackknife test suggest however that the mean and mode are close to each other. 

There remain further caveats to our results; the completeness of the sample, the effect of starburst SED choice, the potential contribution to the far-infrared emission from an AGN, and the contribution from BAL and radio-loud quasars. These caveats are reviewed in \S\ref{caveats}.

\begin{figure}
\includegraphics[width=0.33\textwidth,angle=-90]{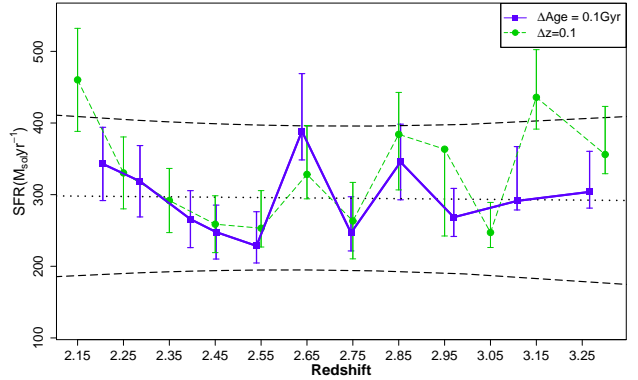} 
\caption{The evolution with redshift of star formation rates in quasar hosts (\S\ref{resredshift}). The data are shown for bins of equal width in both age (blue) and redshift (green). The black lines show the model fit in Eq. \ref{eqnzevo} and the 90\% confidence interval.} 
\label{fig:zslices} 
\end{figure}

\begin{table*} 
\begin{centering} 
\caption{The SPIRE fluxes used to produce Fig. \ref{fig:zslices}. The data for the other plots are given in the supplementary online materials.} 
\small{ 
\begin{tabular}{c|ccccccc}
\hline 
                      & Redshift & 250$\mu$m flux (mJy) & 350$\mu$m flux (mJy) & 500$\mu$m flux (mJy) \\
$\Delta t = 0.1\,$Gyr & 2.20     & 9.12$\pm$0.87        & 11.91$\pm$1.96       & 5.96$\pm$1.78        \\
                      & 2.29     & 7.71$\pm$1.13        & 7.92$\pm$1.16        & 5.71$\pm$1.29        \\
                      & 2.37     & 6.20$\pm$1.04        & 5.89$\pm$1.15        & 3.53$\pm$1.12        \\
                      & 2.45     & 5.84$\pm$1.02        & 5.64$\pm$1.38        & 3.06$\pm$1.19        \\
                      & 2.54     & 4.67$\pm$1.11        & 5.09$\pm$1.20        & 4.33$\pm$1.23        \\
                      & 2.64     & 7.63$\pm$1.37        & 9.98$\pm$1.65        & 4.24$\pm$1.61        \\
                      & 2.75     & 3.63$\pm$1.34        & 6.97$\pm$1.41        & 2.24$\pm$1.43        \\
                      & 2.85     & 5.85$\pm$1.80        & 8.64$\pm$1.83        & 1.12$\pm$1.80        \\
                      & 2.97     & 3.83$\pm$1.66        & 5.85$\pm$1.73        & 2.95$\pm$1.95        \\
                      & 3.11     & 3.80$\pm$1.58        & 5.52$\pm$1.81        & 5.00$\pm$1.78        \\
                      & 3.27     & 3.83$\pm$1.44        & 4.89$\pm$1.44        & 4.24$\pm$1.50        \\ \hline
$\Delta$z = 0.1       & 2.25     & 7.65$\pm$1.25        & 8.27$\pm$1.15        & 4.93$\pm$1.43        \\
                      & 2.35     & 6.68$\pm$0.96        & 5.49$\pm$1.02        & 5.49$\pm$1.10        \\
                      & 2.45     & 6.35$\pm$1.06        & 5.75$\pm$1.37        & 2.65$\pm$1.31        \\
                      & 2.55     & 5.05$\pm$1.48        & 5.94$\pm$1.43        & 4.51$\pm$1.28        \\
                      & 2.65     & 6.64$\pm$1.34        & 8.19$\pm$1.63        & 3.42$\pm$1.55        \\
                      & 2.75     & 3.91$\pm$1.33        & 7.33$\pm$1.37        & 2.27$\pm$1.45        \\
                      & 2.85     & 6.69$\pm$1.78        & 9.14$\pm$1.75        & 1.11$\pm$1.88        \\
                      & 2.95     & 0.10$\pm$2.54        & 3.07$\pm$2.24        & 5.04$\pm$2.60        \\
                      & 3.05     & 2.67$\pm$2.54        & 5.46$\pm$2.47        & 3.24$\pm$2.52        \\
                      & 3.15     & 6.60$\pm$2.05        & 8.31$\pm$2.34        & 4.19$\pm$1.98        \\
                      & 3.28     & 4.68$\pm$1.75        & 4.28$\pm$1.96        & 5.71$\pm$1.57        \\   
\hline 
\end{tabular}
}
\label{tab:redshift} 
\end{centering} 
\end{table*}

\begin{figure}
\includegraphics[width=0.33\textwidth,angle=-90]{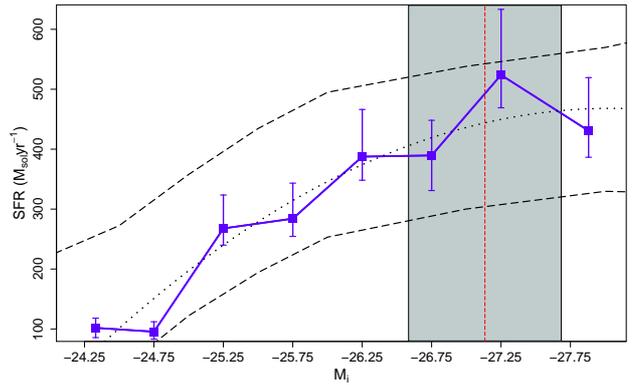} 
\caption{Star formation rate as a function of $M_i$ (\S\ref{resabscol}). The blue points show the data and the black lines show the model in Eq. \ref{eqnsfrimag} and its 90\% confidence interval. The vertical red line and grey shaded region show the $M_i$ and uncertainty for $L^*$ quasars at $2.5 < z < 3.8$ (\citealt{Delvecchio2014}, with K-correction to $z=2$ from \citealt{Richards2006b}.} 
\label{fig:SFR_IMag} 
\end{figure}

\section{Results}\label{sectres}
This section presents the correlations between star formation rates and catalogue quantities given in \citet{Paris2012}. The relations with quantities derived from these quantities are presented in the discussion. The star formation rates can be converted to infrared luminosities via equation 4 of \citet{Kennicutt98}. The flux densities used to infer the results are tabulated; in Table \ref{tab:redshift} for the results in \S\ref{resredshift} and in the supplementary online material for the remainder.

\subsection{Redshift}\label{resredshift}
The evolution of $\dot{\rm{M}}_s$ with redshift is shown in Fig. \ref{fig:zslices}. There is significant scatter, but the data are consistent with an approximately constant mean star formation rate across $2.15<z<3.5$ of $\sim300~\rm{M}_{\odot}$ yr$^{-1}$. This result is independent of whether the data are binned by redshift or lookback time. Fitting a linear model to the $\Delta t$ points gives:

\begin{equation}\label{eqnzevo}
\dot{\rm{M}}_s = (-5\pm 48)z + (309\pm 130) 
\end{equation}

\noindent where $\dot{\rm{M}}_s$ is in units of $\rm{M}_{\odot}$ yr$^{-1}$. This is consistent to within 1$\sigma$ with no evolution. For this model to reproduce the data, the 90\% confidence interval is $\pm100~\rm{M}_{\odot}$yr$^{-1}$. The bins in Fig. \ref{fig:zslices} correspond approximately to $\Delta t=100\,$Myr, which is of order the length of the quasar duty cycle \citep{kelly2010}. However, since we are averaging together $>20$ quasars, each of which represents a (probably) random point in that duty cycle, the bins should be uncorrelated. 

Considering smaller ranges in redshift; the group of five bins spanning $2.15<z<2.55$ shows a decrease in star formation rate from bin to bin, of about $200~\rm{M}_{\odot}\,$ yr$^{-1}$ in total. Moreover, the rise appears statistically significant; the Pearson correlation coefficient {\itshape for just these five points} is $-0.99$. The rise is unlikely to be an artifact of the binning, since we see the same trend in bins of equal $\Delta z$.

\subsection{Absolute Magnitude}\label{resabscol}
Fig. \ref{fig:SFR_IMag} shows the relation between $\dot{\rm{M}}_s$ and $M_i$ (recall from  \S\ref{ssectstack} that $M_i$ samples rest-frame 2500\AA\,). There is a rise in star formation rate with increasing $M_i$. Fitting, purely as a phenomenological choice, a quadratic model yields:

\begin{equation}\label{eqnsfrimag}
\dot{\rm{M}}_s = -(29\pm16)M^2_i - (1638\pm825)M_i - (22505\pm10744)
\end{equation}

\noindent The flattening, or possible decline in star formation rate in the highest $M_i$ bin coincides approximately with the $M_i$ for $L^*$ quasars at $z\sim3$ \citep{Delvecchio2014}. However, the decline is not statistically significant.

\begin{figure}
\includegraphics[width=0.33\textwidth,angle=-90]{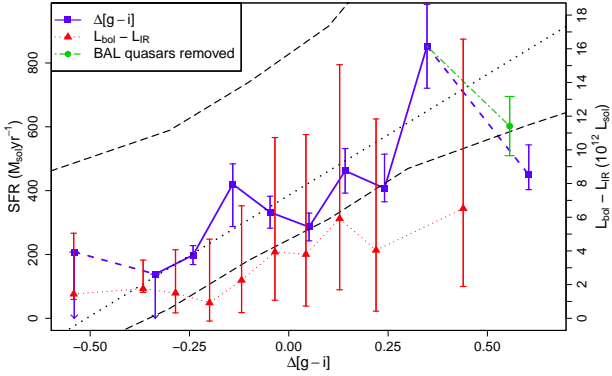}
\caption{Star formation rate vs. $\Delta [g-i]$ (\S\ref{resabscol2}). The blue points show the data. The solid blue line shows the range in $\Delta [g-i]$ over which we have enough objects to be confident of the results (\S\ref{ssectstack}). The black lines show the model in Eq. \ref{eqnsfrrelcol} and its 90\% confidence interval (\S\ref{resabscol}). The points connected by dashed blue lines are those for which we have only a few objects, and are thus less trustworthy. The green point is the same as the last blue point but with the BAL quasars removed. The red line shows the difference between the bolometric and infrared luminosities as a function of $\Delta [g-i]$, according to the right-hand scale (\S\ref{disclum}).} 
\label{fig:SFR_color} 
\end{figure}

\subsection{Colour}\label{resabscol2}
Fig. \ref{fig:SFR_color} shows the relation between $\dot{\rm{M}}_s$ and $\Delta[g-i]$. The distribution of $\Delta[g-i]$ values means we have sufficient objects to investigate this relation only over the range $-0.35 < \Delta[g-i] < 0.35$. We see increasing star formation rates with increasing $\Delta[g-i]$. Fitting a linear model yields:

\begin{equation}\label{eqnsfrrelcol}
\dot{\rm{M}}_s = (758\pm 213)\Delta[g-i] + (385\pm47)
\end{equation}

\noindent We also explored the relation between $\dot{\rm{M}}_s$ and $\Delta[g-i]$ outside the range $-0.35 < \Delta[g-i] < 0.35$. The bins in question have only a few objects so we do not include them in the fit. They are consistent with the $\dot{\rm{M}}_s - \Delta[g-i]$ relation remaining flat at low $\Delta[g-i]$ and turning over at high $\Delta[g-i]$. 

We checked for four possible contributions to these trends. First, we examined the relation between redshift and $\Delta[g-i]$. We found no clear relationship, suggesting that the trends in Fig. \ref{fig:SFR_color} are not dominated by emission lines moving into and out of the $g$ and $i$ bandpasses. Second, we examined how $\Delta[g-i]$ varied with redshift {\itshape within} each bin. Again we found no clear trends; each bin in $\Delta[g-i]$ has approximately the same mean and median redshift. Third, we tested to see if BAL quasars could be affecting the results in any of the bins. The only bin with a significant number of BAL quasars is the last bin. We removed the BAL quasars from this bin and restacked, obtaining the green point in Fig. \ref{fig:SFR_color}. This point is consistent with the original, suggesting that BAL quasars are not dominating our results. Finally, we examined the \textit{difference} in bolometric and infrared luminosity as a function of $\Delta[g-i]$, shown by the red line in Fig. \ref{fig:SFR_color}. This comparison is discussed in \S\ref{disclum2}.

\begin{figure}
\includegraphics[width=0.33\textwidth,angle=-90]{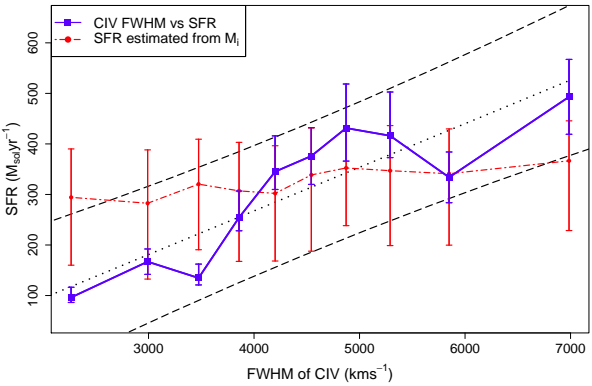} 
\caption{Star formation rate vs. \ion{C}{iv} FWHM (\S\ref{resciv}). The blue points show the data, while the black lines shows the model in Eq. \ref{eqnsfrfwhm} and the 90\% confidence interval. The red line is the result of using Eq. \ref{eqnsfrimag} to calculate the {\itshape predicted} star formation rate associated with the mean $M_i$ for each bin.} 
\label{fig:SFR_FWHMCIV} 
\end{figure}

\subsection{Emission Line Properties}\label{resciv}
The redshifts of our sample means that SDSS catalogue measurements of \ion{C}{iv} are available for nearly all objects (958/1002). Moreover, the FWHM ($F_{\rm{C}}$), EW ($E_{\rm{C}}$), and asymmetry ($A_{\rm{C}}$, see also \S\ref{ssectstack}) of \ion{C}{iv} can be related to physical properties of the AGN. The FWHM scales with black hole mass ($\rm{M}_{bh}$) and rest-frame continuum ultraviolet luminosity ($L_{UV}$) as:

\begin{equation}\label{mbhscalerel}
\rm{M}_{bh} \propto F_{\rm{C}}^{2} L_{UV}^{0.5}
\end{equation}

\noindent \citep{Vestergaard2006,Baskin2005,Sulentic2007}, the EW scales with UV continuum luminosity as:

\begin{equation}
E_{\rm{C}} \propto L_{UV}^{-0.8}
\end{equation}

\noindent \citep{Baldwin1977,kin90}, while $A_{\rm{C}}$ may signpost the presence of AGN winds or other large-scale non-virial motions. We therefore compare star formation rates to these \ion{C}{iv} properties.

First, we examine the $\dot{\rm{M}}_s - F_{\rm{C}}$ relation (Fig. \ref{fig:SFR_FWHMCIV}). We see a positive relationship. A linear model gives a good fit:

\begin{equation}\label{eqnsfrfwhm}
\dot{M}_s = (0.086\pm0.016)F_C - (78\pm73)
\end{equation}

\noindent though the data are also consistent with a power-law with index fixed at 0.5:

\begin{equation}\label{eqnsfrfwhm2}
\dot{M}_s = (-455 \pm 129)F_C^{0.5} + (12\pm2)
\end{equation}

\noindent and if the intercept and slope vary freely then any power-law index below about 1.5 fits the data adequately. 

We performed two tests to see if this rise is driven by a rise in $M_i$ (recalling that $M_i$ samples rest-frame 2500\AA). First, we compared the $F_{\rm{C}}$ and $M_i$ values of our sample, but found no relation. Second, we used Eq. \ref{eqnsfrimag} to convert the average $M_i$ for each $F_{\rm{C}}$ bin into an $M_i$-based star formation rate. These rates are shown by the red line in Fig. \ref{fig:SFR_FWHMCIV}. We still observe an increase in star formation rate, but the trend is shallower. Fitting a linear model to the red line and comparing to that obtained from the fit to the blue line (i.e. assuming that both relations are linear and comparing their distributions of slopes) reveals that the slopes are different at over $4\sigma$ significance. Thus, it is plausible that the increase in $\dot{\rm{M}}_s$ with $F_{\rm{C}}$ is not (primarily) driven by the 2500\AA\, luminosity of the quasar, or vice versa. 

\begin{figure}
\includegraphics[width=0.33\textwidth,angle=-90]{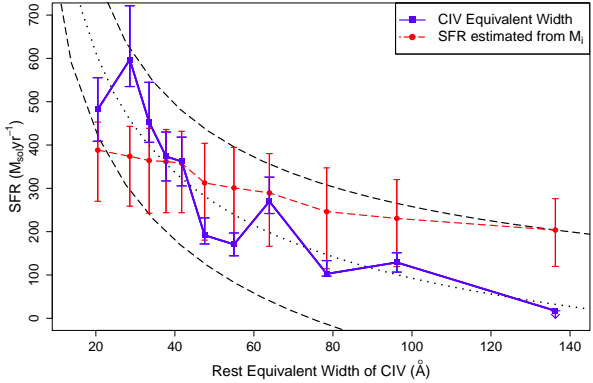} 
\caption{Star formation rate vs. \ion{C}{iv} EW (\S\ref{resciv}). The blue points show the data, while the black lines show the model in Eq. \ref{eqnsfrew} and the 90\% confidence interval. The red line is the result of using Eq. \ref{eqnsfrimag} to calculate the {\itshape predicted} star formation rate associated with the mean $M_i$ for each bin.}\label{fig:SFR_REWECIV} 
\end{figure}

Next, we examine the $\dot{\rm{M}}_s - E_{\rm{C}}$ relation (Fig. \ref{fig:SFR_REWECIV}). We see a decline in star formation rate as $E_{\rm{C}}$ increases. A power-law model with index $-0.5$ fits the data well:

\begin{equation}\label{eqnsfrew}
\dot{M}_s = (4205\pm651)E_{\rm{C}}^{-0.5} - (328\pm98)
\end{equation}

\noindent but the constraints on the index are only that it must be $\gtrsim -1.8$. We again checked to see if this trend could be explained by the dependence on $M_i$, by using Eq. \ref{eqnsfrimag} to convert the average $M_i$ for each $E_{\rm{C}}$ bin into an $M_i$-based star formation rate. These rates are shown by the red line in Fig. \ref{fig:SFR_REWECIV}. The $M_i$ based trend has a flatter slope at well over $4\sigma$ significance. It is thus plausible that the relation between $\dot{\rm{M}}_s$ and  $E_{\rm{C}}$ is not driven solely by the 2500\AA\, luminosity of the quasar.

\begin{figure}
\includegraphics[width=0.33\textwidth,angle=-90]{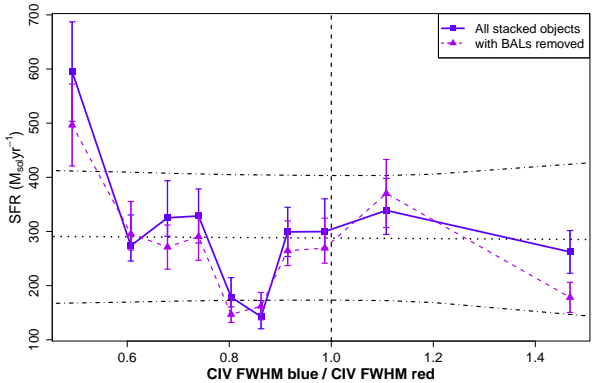} 
\caption{Star formation rate vs. \ion{C}{iv} asymmetry, measured via the blue over red side $E_{\rm{C}}$ (\S\ref{resciv}). The blue points show the data, while the black lines show the model in Eq. \ref{eqncivasym} and the 90\% confidence intervals. The purple line reproduces the blue line, but with the BAL quasars removed.}\label{fig:SFR_blueredCIV} 
\end{figure}

Finally, we examine star formation rate as a function of $A_{\rm{C}}$ (Fig. \ref{fig:SFR_blueredCIV}). Most points are consistent with a flat relation. There are however two bins, $A_{\rm{C}}=0.49$ and  $A_{\rm{C}}=0.86$, that may deviate from a flat relation. Excluding these bins, and fitting a linear model, yields:

\begin{equation}\label{eqncivasym}
\dot{\rm{M}}_s = (5\pm76)A_{\rm{C}} + (293\pm72)
\end{equation}

\noindent This fit is consistent with a flat relation, with a mean close to that of the $\dot{\rm{M}}_s - z$ relation. The deviations of the $A_{\rm{C}}=0.49$ and $A_{\rm{C}}\simeq0.85$ bins from this relation are significant, but only barely so. We assessed the impact of BAL quasars on these trends by excluding them from the stacking. This gives the purple line in Fig. \ref{fig:SFR_blueredCIV}. This line is qualitatively identical to the blue line, suggesting that BAL quasars are not the origin of these results.

\section{Caveats}\label{caveats}

\subsection{Completeness}\label{caveatsoi}
Our sample is the CORE quasar sample from SDSS DR9, and thus is uniformly selected, with a well-understood selection function. In this sense it is among the best available $z>2.2$ quasar samples. It is however not complete. At redshifts close to z=2.2, quasar selection is aided by the strong UV excess of quasars, while at $z>2.2$ the presence of the Ly$\alpha$ forest in the BOSS spectrograph bandpass also helps quasars stand out clearly from stars. At around $z=2.7$ however, quasar colours in the SDSS filters become harder to separate from those of A-type stars, meaning that the completeness at these redshifts is lower.

The completeness of SDSS DR9Q is discussed in \citealt{Ross2012}, their figure 14. They find that the completeness from single-epoch data varies from 70\% at $z=2.2$ to 40\% at $z=2.7$. However, our sample is entirely within Stripe 82. The deeper, multi-epoch data in this field means that the completeness in this field, and thus our sample, should be higher than that in figure 14 of \citealt{Ross2012} (see also \citealt{palde11,Paris2012,mcgreer13}). Nevertheless, completeness in our sample will vary as a function of redshift, and will be lower at $z=2.7$ than at $z=2.2$. We think it unlikely, though not impossible, that completeness could affect either the global flat trend we find, or the apparent upturn at lower redshifts. It is more plausible that some of the 1-2$\sigma$ `structure' in Fig. \ref{fig:zslices} could arise from completeness effects, but we lack the data to assess this possibility. A related issue, the potentially varying contribution from BAL quasars as a function of redshift, is discussed in \S\ref{caveatsiii}.

\subsection{Choice of starburst models}\label{caveatsi}
\begin{figure}
\includegraphics[width=0.33\textwidth,angle=-90]{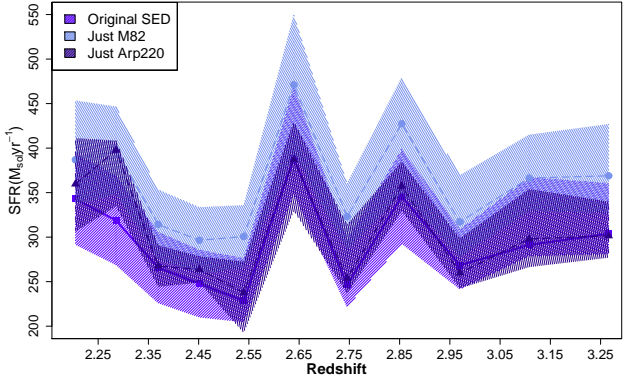} 
\caption{The effect of SED library choice on the $\dot{\rm{M}}_s - z$ relation. This plot reproduces the $\Delta t$ binned data in Fig. \ref{fig:zslices} (medium blue), and adds the same relation derived using more restricted SED libraries; ones that closely resemble M82 (light blue) and Arp220 (dark blue).} 
\label{fig:sedlibrary} 
\end{figure}

We infer star formation rates by fitting starburst models spanning a broad range in parameters. We do so because we have no \textit{a priori} knowledge of the star formation. Nevertheless, this approach is atypical. Most other studies use either a modified blackbody model, or smaller libraries of SEDs that resemble either M82 or Arp220. We thus explore the effect on our results by adopting a more limited model set. In Fig. \ref{fig:sedlibrary} we reproduce the $\dot{\rm{M}}_s - z$ relation in Fig. \ref{fig:zslices} and add the relations obtained using limited libraries corresponding closely to the shape of M82 and Arp220. The M82 library gives star formation rates that are approximately 20\% higher than the original values, while the Arp220 library gives star formation rates that are comparable to the original values, albeit with a larger error. In both cases however the form of the $\dot{\rm{M}}_s - z$ relation is consistent. A similar investigation for the other parameters in \S\ref{sectres} yields similar results. We conclude that our results are not significantly altered by choice of SED library. We regard our errors as more reliable than those computed using a limited model set, since they include uncertainty arising from lack of knowledge of the mode of star formation.

\subsection{Far-infrared emission from AGN}\label{caveatsii}
There is controversy over how much far-infrared emission a `pure' AGN (that is, a system whose IR emission is dominated by dust heated by an AGN, with no significant contribution from dust heated by young or main-sequence stars) can produce. Radiative transfer models for dust around AGN generally predict that the rest-frame far-infrared flux is, for the same total infrared luminosity, factors of at least several lower than the far-infrared flux from a starburst \citep[e.g.][]{Fritz2006,Netzer2007,Mullaney2011}. However, such AGN models usually do not include kpc-scale dust distributions around the AGN. Such a dust distribution could in principle produce substantially greater far-infrared emission. We thus examined the possibility of an AGN contribution to the \textit{Herschel} data via four approaches. 

\begin{figure}
\begin{tabular}{c}
  \includegraphics[width=0.45\textwidth]{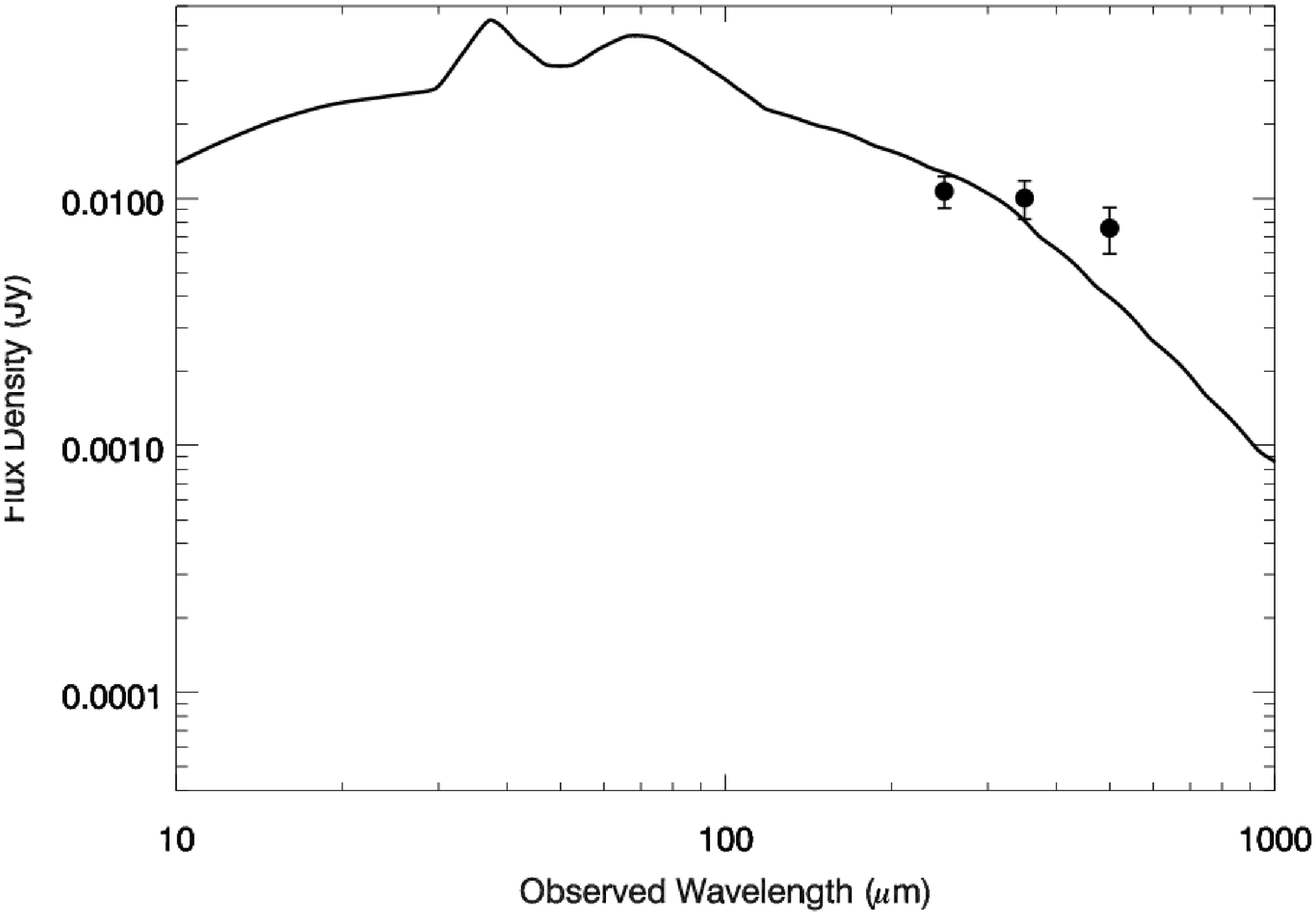} \\
  \includegraphics[width=0.45\textwidth]{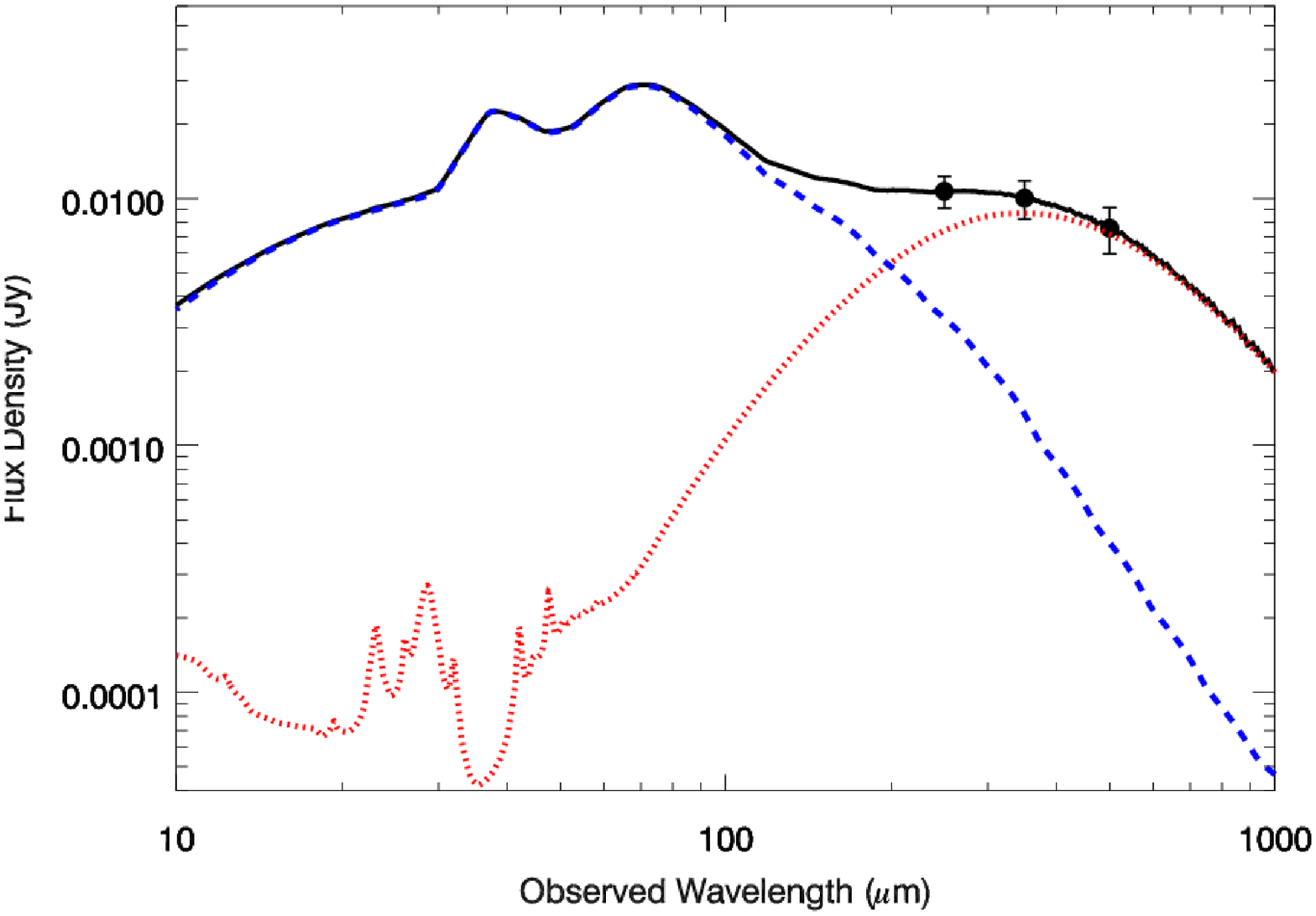} 
\end{tabular}
\caption{Examples of including AGN models when fitting to the SPIRE data. The top panel shows the best-fit pure AGN model to the same SPIRE flux densities as in Fig. \ref{fig:expuresb}. The model can reproduce the 250$\mu$m and 350$\mu$m data but misses the $500\mu$m data. The lower panel shows the result of simultaneously fitting both AGN (blue) and starburst (red) models to the data. The fit is under constrained but still predicts that the starburst contributes most of the SPIRE emission.}
\label{fig:whynoagn} 
\end{figure}

First, we fitted the \textit{Herschel} flux densities with a library of radiative transfer torus models for AGN. We used a different model set to those used in the above-mentioned studies; the library of \citet{Efstathiou1995}. We also imposed the condition that the broad-line region be visible in direct light (in other words, that the viewing angle measured from equatorial is greater than the torus opening angle). As may be expected, the pure AGN fits are in almost all cases both formally rejected, and much worse than the pure starburst fits. We also attempted to fit both the AGN and starburst libraries to the \textit{Herschel} data simultaneously; while the results are highly under-constrained, in all cases a starburst-dominated fit is the best, or joint-best, fit to the data. Examples of both fits are shown in Fig. \ref{fig:whynoagn}. In all cases where a stack has a detection in all three bands, it is the {\itshape shape} of the Herschel fluxes that demands a starburst; an AGN model can explain the 250$\mu$m emission but then not the 350$\mu$m or 500$\mu$m emission. 

Second, we fitted the same library of torus models to the SDSS $griz$ data, and used the best fits to extract predicted \textit{Herschel} flux densities. The results from this test were conceptually the same as from the above test. The AGN models can adequately reproduce the $griz$ data but in doing so nearly always predicted \textit{Herschel} flux densities well below those in the stacks. Those few models that could explain most of the 250$\mu$m emission always fell well below the 350$\mu$m and 500$\mu$m emission. 

Third, we tried fitting AGN and starburst models simultaneously to the SPIRE, WISE and SDSS data in each stack. Since AGN will contribute significantly to the WISE and SDSS data, this gives an alternative way to constrain the AGN contribution to the SPIRE data. However, this approach dramatically increased the complexity of the fits but with no significant increase in the accuracy of either the starburst luminosities, or the far-IR contribution from AGN.

\begin{figure}
\includegraphics[width=0.33\textwidth,angle=-90]{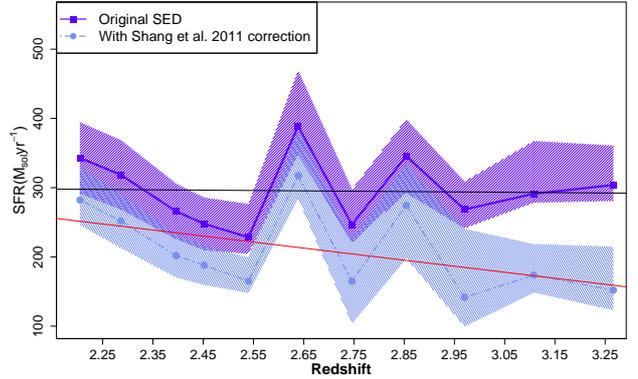} 
\caption{The effect of using the composite observed quasar SED of  \citealt{Shang2011} to correct the $\dot{\rm{M}}_s - z$ relation in Fig. \ref{fig:zslices}. The dark blue region shows the original relation while the light blue region shows the relation after normalizing the \citealt{Shang2011} SED to the median $M_i$ of that bin, extracting predicted SPIRE fluxes, and removing them before fitting. As discussed in \S\ref{caveatsii} this as an overcorrection, so the light blue region should be regarded as a lower limit. Fitting a linear model to the light blue points gives a slight negative slope, but consistent to within 2$\sigma$ with a flat relation.} 
\label{fig:compsedcorr} 
\end{figure}

The above tests do not include extremely spatially extended AGN-heated dust. We do not have models with such distributions, and we lack observations that can disentangle the far-infrared emission on sub-kpc scales. So, as a final test we take a different approach, the use of a composite {\itshape observed} quasar SED to predict \textit{Herschel} flux densities. Composite quasar SEDs are brighter in the far-infrared than models for dusty torii around AGN \citep{elvis94,Richards2006b,Shang2011}. Conversely, in all cases composite SEDs are sparsely sampled at rest-frame $\geq70\mu$m and use far-infrared data of sufficiently coarse resolution that they will include emission from star formation in the quasar hosts; the \citealt{elvis94} sample spans $0<z<0.5$ and uses {\itshape IRAS} data, while the \citealt{Richards2006b} and \citealt{Shang2011} sample have only {\itshape Spitzer} \citep{Werner2004} MIPS 160$\mu$m data. Hence, a composite quasar SED will overcorrect - it will remove both any AGN contribution and the mean star formation rate in the hosts. Nevertheless, we took the composite SED of \citealt{Shang2011}, normalized it to the $M_i$ values for each quasar individually, extracted predicted \textit{Herschel} flux densities, and used these to construct a `corrected' version of Fig. \ref{fig:zslices}. The result is shown in Fig. \ref{fig:compsedcorr}. This plot shows a systematically lower star formation rate, by $\sim30$\%, and a slight negative slope, but the relation between redshift and star formation rate is still consistent with a flat relation. Constructing versions of the other plots with this correction applied reveals similar behaviour - the star formation rates are lower by $\sim30\%$ but the forms of the relations are consistent with our original findings. We deduce that the {\itshape maximum possible} downward correction to the star formation rates due to AGN contamination is approximately 30\%, and that the forms of the relations we find are unlikely to change as a result. 

Overall, none of our tests reveal evidence for a significant AGN contribution to the \textit{Herschel} fluxes. We cannot, however, rule out such a contribution, as all of the tests we can perform have issues; those based on models may miss extended dust, and those based on observed SEDs likely oversubtract due to host galaxy star formation in those SEDs. However, {\itshape the balance of evidence} suggests that rest-frame emission at $\geq70\mu$m from quasars is dominated by star formation, so we interpret our results under the assumption that the \textit{Herschel} data arise purely from star formation.

\subsection{Radio-loud and BAL quasars}\label{caveatsiii}
A small fraction (8.3\%) of our sample are BAL quasars, mostly High Ionization BAL (HiBAL) quasars, which we have treated identically to the non-BAL quasars. In doing so we assume that HiBALs have indistinguishable far-infrared properties from the general quasar population, which is reasonable based on previous results \citep{pri07,gall07,pit15}. An even smaller fraction (1.9\%) have radio data, and it is likely that many of these objects are radio-loud. However, we have not excluded these objects from our samples. We here explore the effects of these decisions.

\begin{figure}
\includegraphics[width=0.33\textwidth,angle=-90]{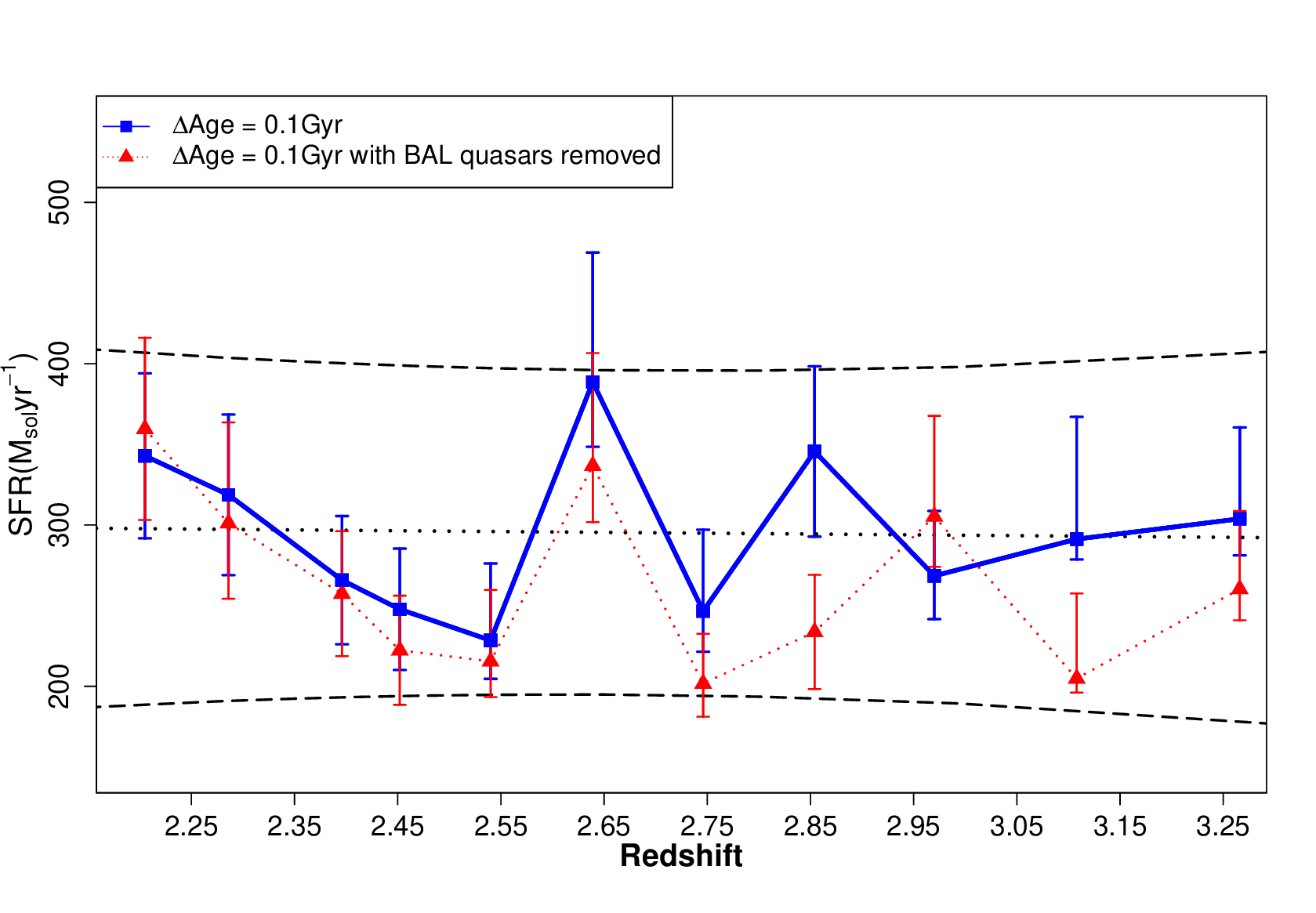} 
\includegraphics[width=0.33\textwidth,angle=-90]{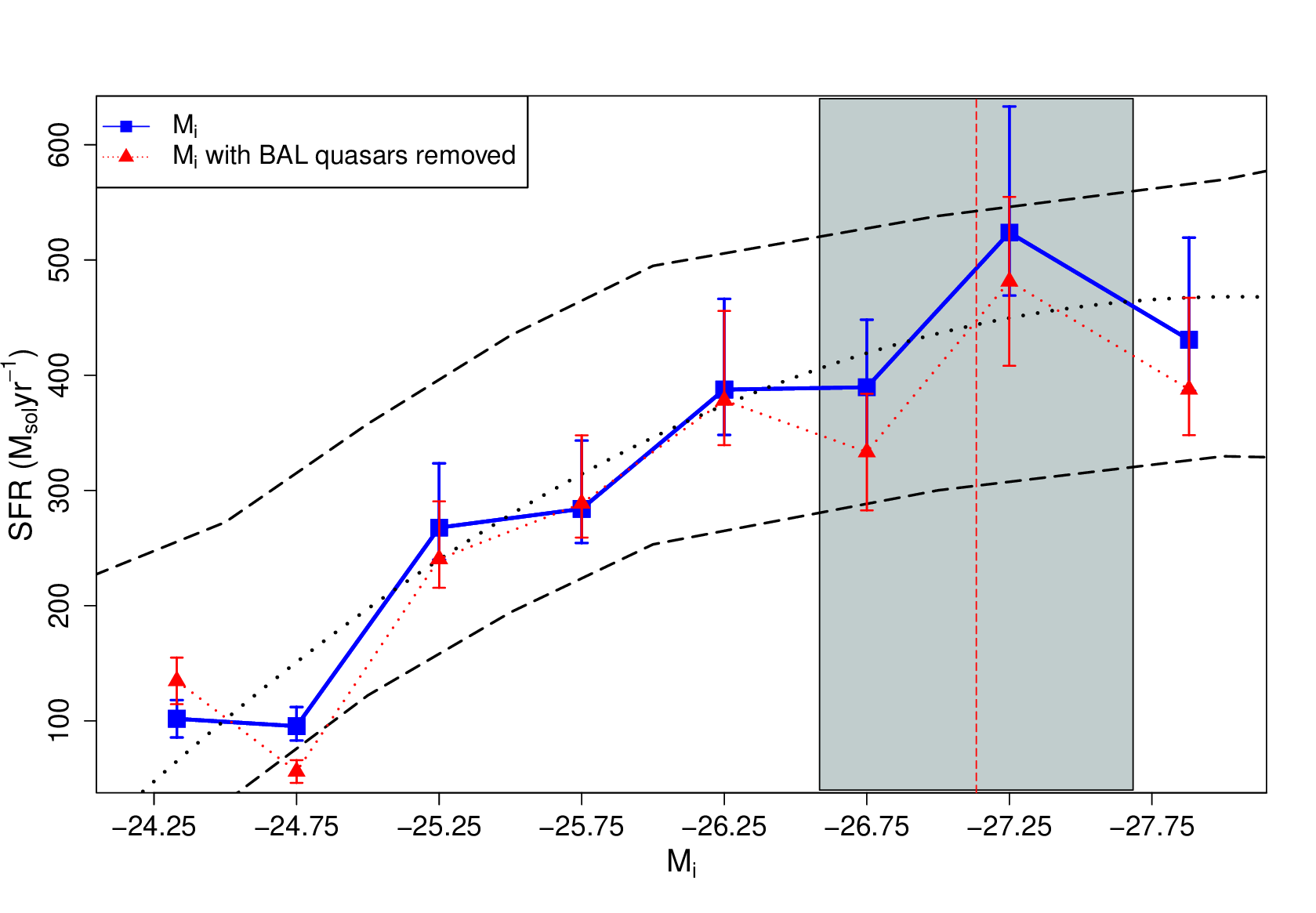} 
\caption{The effect on Figs. \ref{fig:zslices} and \ref{fig:SFR_IMag} if we remove BAL quasars from the stacks. In neither case do we see significant differences, despite the varying fraction of BAL quasars in each bin.} 
\label{fig:zevonobalcomp} 
\end{figure}

Both the BAL quasars and the quasars with radio data have identical (within the errors) distributions in redshift and absolute magnitude as the rest of the sample. However, previous papers have shown that both BAL and radio-loud quasars differ in other respects from classical quasars. The radio detected quasars comprise less than 4\% of the quasars in virtually all the stacks. Even if all these quasars are radio-loud, it is unlikely that their inclusion has a significant effect. The BAL quasars on the other hand, while comprising less than 10\% of the quasars in most stacks, are in a few cases over 15\%. We thus investigated the impact of BAL quasars on our results by repeating the stacking analyses with the BAL quasars removed. Two examples, for redshift and $M_i$, are shown in Fig. \ref{fig:zevonobalcomp}. A further example is shown in Fig. \ref{fig:SFR_color}. In no case did we find any significant differences, either in terms of mean values or shapes of trends. We conclude that neither the BAL or radio-loud objects in our sample are significantly affecting the trends we observe.

\section{Discussion}
Our results provide insight on both the evolution with redshift of star formation rates in quasar host galaxies, and on the relationship between star formation and AGN activity, at $2.15<z<3.5$. By stacking \textit{Herschel} SPIRE photometry for a sample of 1002 optically selected quasars, we examine {\itshape typical} star formation rates in high redshift, optically luminous, unobscured AGN, rather than individually detected objects that represent the more extreme star formation events. However, because we require a minimum of 20 objects per bin to achieve robust detections, we can investigate at most 10-12 bins in total, so we cannot reliably explore degeneracies in relationships between parameters. Moreover, when discussing relationships between star formation rate and \ion{C}{iv} line properties, we assume that the \ion{C}{iv} line arises exclusively from the AGN.

\subsection{Redshift}\label{cosmoevo}
We find no evidence for strong evolution of star formation rates in quasar hosts with redshift across $2.15<z<3.5$. Instead, we find an approximately constant mean star formation rate of $300\pm100~\rm{M}_{\odot}$yr$^{-1}$. Our mean rate is higher than the star formation rates seen in $z\lesssim2$ AGN \citep{Lacy2007,Silver09,Floyd2013,Huseman2014,ban15}, but lower than those found in far-infrared luminous (i.e. individually detected) quasars at similar redshifts \citep{lut08,pit15}. Finally, if the stellar masses of the host galaxies are of order $10^{11}$M$_{\odot}$ then our mean rate lies on or somewhat above the `main sequence' star formation rate at $z\sim2$ \citep{Elbaz2011,Magnelli2013,Gruppioni2013}

The simplest interpretation of this result, together with the relatively flat comoving star formation and quasar luminosity densities over $2<z<3.5$, is that the processes that trigger quasars evolve in a similar way to the processes that trigger star formation.  Furthermore, it is consistent with the bulk of star formation in quasar hosts at this epoch not being in the `starburst' mode \citep{rod11}. However, further interpretation depends on the quasar host masses. There is evidence that, at $z>1$, the  $\rm{M}_{bh}/\rm{M}_s$ ratio is higher than at low redshift (that is, the hosts of high redshift quasars are less massive than low redshift quasars for the same black hole mass), though the redshift {\itshape and} luminosity evolution of this ratio remain uncertain \citep{peng06,shie06,salv07,woo08}. There also exists a wide range in stellar mass for a given star formation rate at $z>1$; our mean star formation rate spans a range of over 1 dex in stellar mass \citep{Wuyts2011}. Since we have no measures of the host galaxy masses, we cannot make further inferences from the flat star formation rate relation. 

The factor of $\sim2$ rise in $\dot{\rm{M}}_s$ from $z=2.5$ to $z=2.1$ is not straightforward to explain. This $\Delta z$ corresponds to $\Delta t=0.5$Gyr, a factor of a few longer than the quasar duty cycle \citep{kelly2010}. However, the (rest-frame ultraviolet) galaxy luminosity function does not evolve substantially over $2<z<3$ \citep{Arn2005,Red2009,Ha2010,Kho2015}. Using a Schechter function then $\phi^{*}$ and $\alpha$ change little over this redshift range, while $\rm{M}^{*}$ may change somewhat more \citep{Parsa2015}. A factor of $\sim2$ change in $\dot{\rm{M}}_s$ over $2.1<z<2.5$ could thus plausibly have a physical origin, related to the masses of galaxies in which quasar activity is triggered. It is however also plausible that this rise is due to an unaccounted for effect in the selection of the DR9 quasar catalog.

\subsection{Black Hole accretion rate}\label{disclum}
The $M_i$ values in Fig. \ref{fig:SFR_IMag} can be converted into a bolometric accretion luminosity $L_{b}$ via:

\begin{equation}\label{lbolmi}
\frac{L_{b}}{L_{\odot}} = \kappa\left(3.72\times10^{21}\right)10^{-0.4[M_i + 49.79]} 
\end{equation}

\begin{figure}
\includegraphics[width=0.33\textwidth,angle=-90]{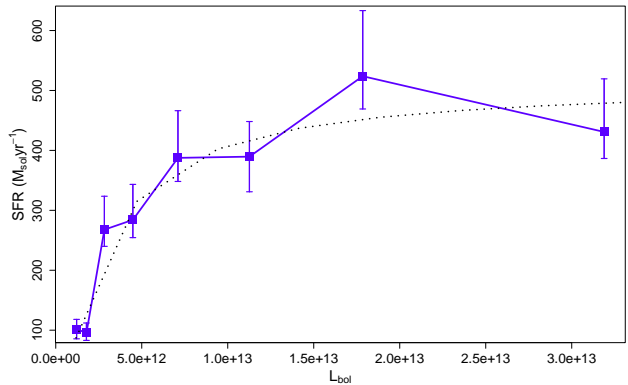} 
\caption{Star formation rate as a function of bolometric accretion luminosity. The $L_{b}$ values were computed via Eq. \ref{lbolmi}. The dotted line is the relation in Eq. \ref{lbolsfr}.} 
\label{fig:SFR_lbol} 
\end{figure}

\noindent (see also \citealt{Richards2006}, their equation 4) where $\kappa\simeq5$ is the correction from $\nu L_{\nu}$ at 2500\AA. This leads to the $\dot{\rm{M}}_s - L_{b}$ relation shown in Fig. \ref{fig:SFR_lbol}. An unwieldy expression reproducing this relation can be obtained by substituting Eq. \ref{lbolmi} into Eq. \ref{eqnsfrimag}, but a more palatable expression that almost exactly reproduces it is:

\begin{equation}\label{lbolsfr}
\dot{\rm{M}}_s = (11\pm135) + (504\pm114) exp\left[\frac{(2.4\pm1.5)\times10^{12}}{L_{b}}\right]^{-1}
\end{equation}

\noindent Assuming that the relation between $L_{b}$ and $\dot{\rm{M}}_{bh}$ is linear:

\begin{equation}\label{eqnlaccmbh}
L_{b} = \eta \dot{\rm{M}}_{bh}c^2 
\end{equation}

\noindent (where $\rm{M}_{bh}$ is the mass of the black hole and $\eta$ is the fraction of gravitational potential energy radiated away by infalling material, thought to range between 0.06 and 0.42 for a Schwarzschild and Kerr black hole) means that Eq. \ref{lbolsfr} is also the form of the relation between $\dot{\rm{M}}_s$ and $\dot{\rm{M}}_{bh}$.

Fig. \ref{fig:SFR_lbol} and Eq. \ref{lbolsfr} are consistent with the idea that there may be a `maximal' {\itshape typical} star formation rate in optically selected type 1 quasar hosts of $\sim600$M$_{\odot}$ yr$^{-1}$, beyond which star formation rates do not rise with increasing accretion rate. Below this value (i.e. neglecting the last point in Fig. \ref{fig:SFR_lbol}), the points are consistent with models ranging from a linear relation with a zero intercept

\begin{equation}\label{lbolsfrlin}
\frac{\dot{\rm{M}}_s}{\rm{M_{\odot}yr^{-1}}} = (3.56\pm0.50)\times10^{-11} \frac{L_{b}}{L_{\odot}}
\end{equation}

\noindent to a power law with index 0.5. Assuming a linear model, converting to SI units and substituting from Eq. \ref{eqnlaccmbh} yields

\begin{equation}\label{eqndotmsdotmbh}
\frac{\dot{\rm{M}}_{bh}}{\dot{\rm{M}}_{s}} = \frac{1.91\pm0.27}{\eta}\times10^{-3}
\end{equation}

\noindent A maximal star formation rate is consistent with the idea that star formation rates in quasar hosts (at least in the domain examined here) `saturate' at high luminosity, perhaps due to supernova winds (see also \citealt{silk13,gea13}). However, the {\itshape existence} of a correlation below this value is more controversial. Our finding of a correlation is consistent with some previous studies on quasars, obscured AGN and star-forming galaxies, although most of these studies sample lower $L_{b}$ and $z$ ranges \citep[e.g][]{net09,hatz10,ima11,raff11,mull12a,chen13,Young2014,delv15,xu15}. Moreover, the form of the relation we find is consistent with these studies, which also find that it is linear, or close to linear (e.g. \citealt{mull12a} find a linear relation while \citealt{net09,xu15} find $L_{SF}\propto L_{AGN}^{0.8}$). Conversely, other studies find a weak, or no relation between  $\dot{\rm{M}}_s$ and $\dot{\rm{M}}_{bh}$, although again these studies are mostly at lower $L_{b}$ and/or $z$ \citep[e.g.][]{pri03,Shao2010,dic12,Mull2012,harr12,Rosario2013,Stan2015,pit15}

To uncover the origin of this contrast, we consider the ways in which our study differs from previous work; we sample higher AGN luminosities, we have more objects at $z>2$, and we infer relationships via stacking, rather than individual detections. These three differences mean that there are three possible reasons why we see a relationship between $\dot{\rm{M}}_s$ and $\dot{\rm{M}}_{bh}$, while some other studies have not. 

The first is that an $\dot{\rm{M}}_s - \dot{\rm{M}}_{bh}$ relation only emerges at high redshift, around $z=2$ (see also \citealt{hatz10,rov12,delv15}). A motivation for this possibility is that the total free gas to stellar mass fraction, $f_{gas}$, rises with redshift as  $\sim(1+z)^2$ up to at least $z=1$, and may plateau at $z\sim3$ \citep[e.g.][]{ler08,sain11,mag12,tac13,tro14,pop15,lag15}. Other studies have suggested that systems with higher $f_{gas}$ are more likely to have higher star formation rates, and that there is a strong connection between available cold gas and the probability that a black hole is accreting rapidly \citep[e.g.][]{gen10,vit14}. This possibility would explain the emergence of trends in our study, which (some) previous studies did not find. It is also consistent with the idea that both star formation and black hole accretion depend on the availability of free baryons. 

The second is that short-term ($<100\,$Myr) AGN variability introduces scatter in the $\dot{\rm{M}}_s - \dot{\rm{M}}_{bh}$ relation derived from measurements of objects individually. This possibility has been suggested both from simulations \citep{gab13,Volon2015} and observations \citep{hickox14,Stan2015}. Since we stack large numbers of objects together, any reasonable level of AGN variability would be averaged out. This possibility could thus also explain the contrast between our results and previous work. 

The third is that an $\dot{\rm{M}}_s - \dot{\rm{M}}_{bh}$ relation only emerges at high $L_{b}$, around $3\times10^{12}$L$_{\odot}$ (see also \citealt{lut08,Shao2010,Ros12,rov12,ban15}). A motivation for this possibility is that black hole growth may correlate with star formation on $\sim$sub-kpc scales, but not with star formation on $\gtrsim$kpc scales \citep[e.g.][]{DiaRie2012,Volon2015}. If, at high $L_{b}$, star formation in quasar hosts shifted to smaller spatial scales (perhaps because of a higher fraction of merger-triggered rather than secular star formation), then the emergence of a $\dot{\rm{M}}_s - \dot{\rm{M}}_{bh}$ relation in our sample is natural. This explanation is however not wholly satisfactory. It is in tension with the studies of lower luminosity AGN that do find a correlation, and with studies of quasar hosts with extremely high star formation rates that do not find a strong $\dot{\rm{M}}_s - \dot{\rm{M}}_{bh}$ relation \citep{pit15}.

We cannot discriminate between these three possibilities, as to do so requires a larger sample, and X-ray data. We thus speculate, based on our results and previous work, that a correlation between $\dot{\rm{M}}_s$ and $\dot{\rm{M}}_{bh}$ exists in certain parts of the $z - \dot{\rm{M}}_s - \dot{\rm{M}}_{bh}$ parameter space, dependent on the availability of free baryons, the trigger for activity (major mergers, secular processes, etc), the potential positive and negative effects of AGN on star formation. Moreover, the form of the correlation - how strong it is, how non-linear it is - also may vary based on the same factors.

\subsection{Black Hole mass}\label{discmass}
\begin{figure}
\includegraphics[width=0.33\textwidth,angle=-90]{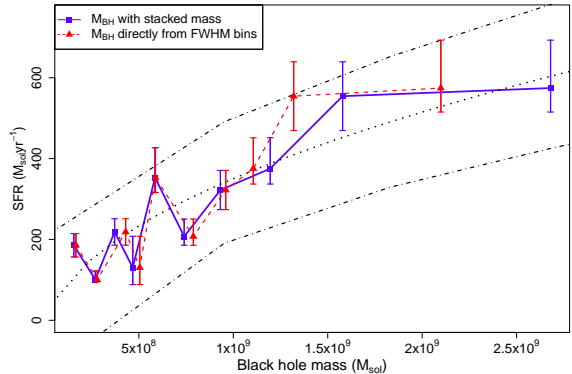} 
\caption{Star formation rate as a function of black hole mass. The results from two approaches are presented; stacking on black hole mass directly, and computing the mean black hole masses for the bins in Fig. \ref{fig:SFR_FWHMCIV}. The results from both are identical within the errors. The black lines show the model in Eq. \ref{eqnsmbhsfr} and the 90\% confidence interval.} 
\label{fig:SFR_smbh} 
\end{figure}

\begin{figure}
\includegraphics[width=0.33\textwidth,angle=-90]{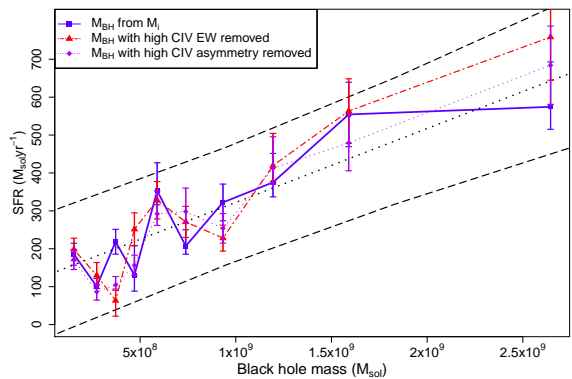} 
\caption{The effect on the relation in Fig. \ref{fig:SFR_smbh} if quasars with high $E_{\rm{C}}$ or high $A_{\rm{C}}$ values are removed from the stacks. In neither case do we see significant differences.} 
\label{fig:SFR_smbh2} 
\end{figure}

Fig. \ref{fig:SFR_FWHMCIV} is consistent with the idea that higher star formation rates are found in quasars with more massive black holes. The data in \citet{Paris2012} preclude an optimal calculation of black hole mass, but the $M_i$ values sample the rest-frame UV (\S\ref{ssectstack} and \citealt{Paris2012}). We thus estimate black hole masses by converting the $M_i$ values to a monochromatic luminosity at 1450\AA\ ($L_{1450}$) using equation 3 of \citet{Richards2006} and then use these luminosities and the $F_{\rm{C}}$ values to compute $\rm{M}_{bh}$ via equation 7 of \citet{Vestergaard2006}. We performed the stacking in two ways; stacking directly on black hole mass, and calculating the mean black hole masses for each bin in Fig. \ref{fig:SFR_FWHMCIV}. The results from both approaches are shown in Fig. \ref{fig:SFR_smbh}, and are identical within the errors. We see a positive correlation. A power-law model fits the data well but the constraints on the index are weak. An index of 0.5 gives an acceptable fit:

\begin{equation}\label{eqnsmbhsfr}
\dot{\rm{M}}_s = \left[ (1.3\pm0.22)\times10^{-2} \right]\rm{M}_{bh}^{0.5} - (47\pm65) 
\end{equation}

\noindent but so does a linear model:

\begin{equation}\label{eqnsmbhsfr2}
\dot{\rm{M}}_s = \left[ (1.92\pm0.35)\times10^{-7} \right] \rm{M}_{bh} + (103 \pm 41) 
\end{equation}

\noindent and a model with an index of $1/5$. As a check on the validity of using the $M_i$ values to compute $L_{1450}$, we estimated $L_{1450}$ from the SDSS $g$-band magnitudes using the same spectral index, and then used these data to compute $\rm{M}_{bh}$. The resulting fit was virtually identical to that obtained by using $M_i$. We also checked the effect on these results if the quasars with high $E_{\rm{C}}$ or high $A_{\rm{C}}$ values were removed from the stacks. For $E_{\rm{C}}$ we removed those objects in the last bin in Fig. \ref{fig:SFR_REWECIV} while for $A_{\rm{C}}$  we removed those objects in the first and last bins in Fig. \ref{fig:SFR_blueredCIV}. The results are shown in Fig \ref{fig:SFR_smbh2}. In neither case did we see significant differences.

It is illuminating to compare this result to the relations between $\dot{\rm{M}}_s$ and both $F_C$ and $\dot{\rm{M}}_{bh}$. From \S\ref{resciv} we can conclude that $\dot{\rm{M}}_s \propto F_C^{\alpha}$, where $0.5 \lesssim \alpha \lesssim 1.4$. From \S\ref{disclum} we can conclude that $\dot{\rm{M}}_s \propto L_b^{\beta}$, where $0.4 \lesssim \beta \lesssim 1.1$. Assuming that Eq. \ref{mbhscalerel} holds then the above findings mean that a relation of the form in Eq. \ref{eqnsmbhsfr} is expected. 

We now address the question; do star formation rates in quasar hosts scale with black hole mass, black hole accretion rate, or both? Given the limitations of our data we can only address this question in a simple way. If $\dot{\rm{M}}_s$ scaled solely with $\dot{\rm{M}}_{bh}$ then we would not expect to see a strong $\dot{\rm{M}}_s - F_{\rm{C}}$ relation, but we clearly do. If however $\dot{\rm{M}}_s$ scaled solely with $\rm{M}_{bh}$, via e.g. $\dot{\rm{M}}_s^{\gamma} \propto \rm{M}_{bh}$, then we would expect $\dot{\rm{M}}_s - F_{\rm{C}}$ and $\dot{\rm{M}}_s - L_b$ relations that were both power laws, which is what we find. Thus, our results are consistent with star formation rates in quasar hosts scaling with black hole mass. We find no evidence that favours an additional (i.e. beyond that implied by Eq. \ref{mbhscalerel}) scaling between $\dot{\rm{M}}_s$ and $\dot{\rm{M}}_{bh}$. 

This is consistent with a common gas reservoir fueling both the growth of the black hole and star formation, such that a larger (or more optimally distributed) reservoir leads to both a larger black hole and higher star formation rates. This scaling relation would manifest on timescales of $100-200$~Myr, comparable to both the lifetime of a starburst, and the quasar duty cycle. We thus cautiously propose that the scaling between $\dot{\rm{M}}_s$ and $\rm{M}_{bh}$ is the most important one for understanding the relationship between black hole mass and stellar mass in quasar hosts. This idea though depends on the starburst `saturating' so as to allow Eq. \ref{lbolsfr} to be interpreted as we do. 

We cannot, however, rule out a separate relation between $\dot{\rm{M}}_s$ and $\dot{\rm{M}}_{bh}$. There is room within the uncertainties on the scaling relations between $\dot{\rm{M}}_s$ and both $F_C$ and $L_b$ such that we can have a relation between $\dot{\rm{M}}_s$ and $L_b$ as implied by a $\dot{\rm{M}}_s^{\gamma} \propto \rm{M}_{bh}$ relation, and an additional dependence on $L_b$. Such a dependence could be driven on timescales comparable to those required for significant change in $\dot{\rm{M}}_{bh}$, of order tens of Myr. These processes could include how efficiently gas is channeled into the $\lesssim \,$kpc regions of the host, thus regulating both $\dot{\rm{M}}_{s}$ and $\dot{\rm{M}}_{bh}$.

Two related points are worth noting. First, some authors have argued that \ion{C}{iv} is a worse tracer of $\rm{M}_{bh}$ at high luminosities than \ion{Mg}{ii} or H$\beta$ \citep{Baskin2005,Netzer2007}. Possible sources of contamination could include the \ion{C}{iv} emission arising from greater distances than the other lines, and/or significant non-virial motions. However, the overall scaling between $F_{\rm{C}}$ and $\rm{M}_{bh}$ has been argued to be reliable \citep{assef11}. Moreover, excluding the objects with highly asymmetric \ion{C}{iv} lines does not appreciably change the relation (Fig. \ref{fig:SFR_smbh2}). We conclude that the \ion{C}{iv} lines, at least for a study like ours which averages over tens of objects, is a reasonable way to obtain black hole masses. Second, there is controversy over the relation between accretion rate and black hole mass; \citet{Netzer2007} find no correlation of $L/L_{e}$ (in which $L_{e}$ is the Eddington luminosity) with $\rm{M}_{bh}$, while \citet{Bonfield2011} do, albeit with a large scatter.

\subsection{AGN Colour}\label{disclum2}
The $\Delta[g-i]$ variable is a crude measure (i.e. using broad filters, and without accounting for the contribution from lines) of the UV continuum slope over $2<z<3$, spanning (1500\AA-1200\AA)/(2500\AA-1900\AA). There is evidence that the UV continua of quasars become bluer with increasing bolometric luminosity (\citealt{saka11,koku14,xie15}, but also \citealt{kraw13}), so the positive correlation of $\Delta[g-i]$ with star formation rate in Fig. \ref{fig:SFR_color} is plausibly a manifestation of the relation with $M_i$ in Fig. \ref{fig:SFR_IMag}. The possible turnover at the most positive $\Delta[g-i]$ values {\itshape could} be a separate, evolutionary effect, namely low star formation rates in slightly dust-reddened quasars in a post-starburst phase. There may also be a rise in $L_{b}-L_{IR}$ with increasing $\Delta[g-i]$ (the red line in Fig. \ref{fig:SFR_color}), which could suggest an AGN origin for part of the infrared emission. We do not however regard this as likely, for the reasons given in \S\ref{measrates} and \S\ref{caveats}. Instead, if this rise is real, we interpret it as a larger fraction of the total output arising from star formation.

\subsection{Eddington Ratio}\label{disceddr}
Assuming opacity via Thompson scattering through Hydrogen, the Eddington ratio, $\lambda_{edd}$, of a quasar is given by:

\begin{equation}\label{eqnlambdaedddef}
\lambda_{edd} = \frac{L_{b}}{L_{e}} \simeq \frac{\epsilon \dot{\rm{M}}_{bh} c^2 \sigma_{T}}{4 \pi  G \rm{M}_{bh} m_{p}}
\end{equation}

\noindent in which $L_{e}$ is the Eddington luminosity, $\sigma_{T}$ is the Thompson scattering cross-section for the electron, and $m_p$ is the mass of the proton. Bolometric luminosities are calculated from Eq. \ref{lbolmi}. This examination has the caveat that the resulting $\lambda_{edd}$ distribution is skewed towards values at $<0.2$, so, to ensure enough objects per bin to obtain reasonable error bars, we place most of the bins at $\lambda_{edd}<0.2$, with only a few at higher values. 

\begin{figure}
\includegraphics[width=0.33\textwidth,angle=-90]{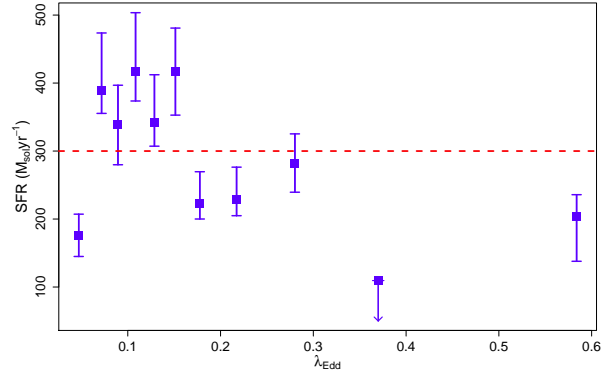} 
\caption{Star formation rate as a function of $\lambda_{edd}$ (Eq. \ref{eqnlambdaedddef}).  The red horizontal line is the mean star formation rate derived from the model fit in Fig. \ref{fig:zslices}.} 
\label{fig:SFR_eddrat} 
\end{figure}

The results are shown in Fig. \ref{fig:SFR_eddrat}. There is, ostensibly, a trend; low $\lambda_{edd}$ values mostly correspond to higher star formation rates than do high $\lambda_{edd}$ values. The uneven, sparse binning means however that we cannot be certain of this, as the data are also consistent with a flat relationship with a mean that is approximately the same as the mean star formation rate with redshift in Fig. \ref{fig:zslices}. 

A lack of any obvious trend between $\dot{\rm{M}}_s$ and $\lambda_{edd}$ is straightforward to understand in the context of the $\dot{\rm{M}}_s - \rm{M}_{bh}$ and $\dot{\rm{M}}_s - \dot{\rm{M}}_{bh}$ relations. If the underlying driver is the relation between star formation rate and black hole mass, with a weaker or no independent relation with accretion rate, then there would no reason to expect a relation between star formation rate and how {\itshape efficiently} the black hole is accreting. 

It is however informative to speculate on the opposite case, that the hint of a trend of low $\lambda_{edd}$ values corresponding to higher star formation rates is real. There are two straightforward interpretations. First, that this is a manifestation of smaller mass black holes accreting more efficiently. Second, that the peak star formation rate occurs some time before or after the peak in the AGN luminosity.

\subsection{AGN Winds}
We find an approximately constant $\dot{\rm{M}}_s$ as a function of A$_{C}$ (Fig. \ref{fig:SFR_blueredCIV} \& Eq. \ref{eqncivasym}). This relation is consistent with the idea that line asymmetries arise due to the relative orientation of the quasar \citep{Richards2002}; assuming that the infrared emission from star formation is optically thin, then variations in line asymmetry would be accompanied by no net variation in star formation rate.

There are two deviations from the model in Eq. \ref{eqncivasym}. First is a dip in  $\dot{\rm{M}}_s$ at $A_{\rm{C}}\simeq0.85$. Second is an enhancement in $\dot{\rm{M}}_s$  at $A_{\rm{C}}\simeq0.5$. Both features are barely significant, however, we explore their implications further. The idea that line asymmetries arise due to the relative orientation of the quasar does not explain either deviation. Instead, we speculate that these features are evidence for the black hole affecting star formation in the host. In this context the asymmetric \ion{C}{iv} emission signposts outflowing gas. The rise at very blue asymmetric values is consistent with the idea that such outflows can {\itshape trigger} starbursts in gas-rich systems (e.g. \citealt{Zubovas2013}). Conversely, the dip at moderately blue asymmetries is consistent with the idea that AGN outflows can quench star formation \citep[e.g.][]{Fabian2012}. The weakness of both effects could be due to the AGN duty cycle being much longer than the timescale for feedback, leading to only a faint signal in a statistical study such as ours. Neither explanation is however wholly satisfactory. Quenching in particular is more commonly associated with BAL winds \citep[e.g.][]{Farrah2012}, and there is no motivation for why we should only see quenching over a certain range in \ion{C}{iv} asymmetry. 

\subsection{The Baldwin Effect}\label{discbald}
We observe a declining $\dot{\rm{M}}_s - E_{\rm{C}}$ relation in Fig. \ref{fig:SFR_REWECIV}, and a rising $\dot{\rm{M}}_s - M_i$ relation in Fig. \ref{fig:SFR_IMag}. At face value, this is consistent with the Baldwin effect. This effect \citep[e.g.][]{Baldwin1977,Wilkes1999,Green2001,Xu2008,wu2009} - a rise in L$_{2500}$ accompanied by a decline in \ion{C}{iv} EW - has a proposed origin in a nonlinear relation between 2500\AA ~luminosity and optical to X-ray spectral slope ($\alpha_{ox}$). In it, a rise in the number of optical photons is met with a relative decrease in the number of X-ray photons that can {\itshape produce} \ion{C}{iv}, meaning that $M_i$ rises as $E_{\rm{C}}$ declines. Models for the Baldwin effect include a quasar SED that is softer at higher luminosities (\citealt{Netzer1992,Dietrich2002}, see also \S\ref{disclum2}), and that ultraviolet luminosity is less isotropic than X-ray luminosity \citep[e.g.][]{Wilkes1999}. There is no evidence that the Baldwin effect evolves with redshift at $z\geq2$ \citep{Osmer1994,Dietrich2002,Xu2008,Bian2012} and controversy over the relation with (\ion{C}{iv} derived) black hole mass \citep{Xu2008,Bian2012}. However, \ion{C}{iv} EW anticorrelates with $\lambda_{edd}$ \citep{Boroson1985,Bachev2004,Warner2004,Baskin2004}, including with Eddington ratios derived from \ion{Mg}{ii} \citep{Bian2012}.

We propose however that the Baldwin effect alone cannot explain the relation in Fig. \ref{fig:SFR_REWECIV}. The dynamic range in $\dot{\rm{M}}_s$ in Fig. \ref{fig:SFR_REWECIV} is greater than the dynamic range in $\dot{\rm{M}}_s$ in Fig. \ref{fig:SFR_lbol}. Moreover, taking the mean $M_i$ for each bin in Fig. \ref{fig:SFR_REWECIV} and then using Eq. \ref{eqnsfrimag} to derive the {\itshape expected} star formation rates for the EW-binned data yields the red line in Fig. \ref{fig:SFR_REWECIV}, a significantly flatter relation. It is likely therefore that additional factors contribute to what we see. 

There are four possible candidates for these additional factors. First is that $M_i$ is not a linear tracer of L$_{2500}$. Second is that the Baldwin effect changes form at high $L_b$. Third is that high $E_C$ values signpost a change in the scaling relation between 
$\dot{\rm{M}}_s$ and $\dot{\rm{M}}_{bh}$ (see also \S\ref{discmass}). Fourth is that this is evidence for quenching of star formation by the AGN; if the EW of \ion{C}{iv} is a proxy for the radiant intensity available to drive winds into the host then large \ion{C}{iv} EWs would signpost the {\itshape population} of quasars in which the AGN could exert maximal influence on star formation.

The first, second, and third candidates seem plausible, but we lack the data to confirm or refute them. The fourth possibility is also, at face value, plausible, since we find lower star formation rates for the highest $E_{\rm{C}}$ values compared to the lowest $M_i$ values. We do not however regard it as a likely contributor. Assuming a two component disk+wind model, then a high X-ray luminosity will suppress a line-driven wind by over-ionizing the gas in the BELR. Furthermore, strongly blueshifted \ion{C}{iv} is associated with weaker X-ray spectra, and quasars with high $E_{\rm{C}}$ and highly blueshifted \ion{C}{iv} are depopulated in SDSS DR9 \citep{Richards2011}. The only way (within the above framework) that quenching could contribute is if the wind is disk-launched, and having the continuum filtered though a more ionized continuum makes a more powerful outflow more likely.

\section{Conclusions}
We have stacked {\itshape Herschel} SPIRE data on the positions of 1002 optically selected type 1 quasars with $-28.6<M_{i}<-23.8$ spanning $2.15<z<3.5$. We used the resulting flux densities at 250, 350 and 500 $\mu$m to infer star formation rates as a function of redshift, black hole accretion rate, black hole mass, Eddington ratio, optical color, and \ion{C}{iv} line asymmetry. Our conclusions are:

1 - Star formation rates in quasar hosts remain approximately constant with redshift across $2.15<z<3.5$, at $300\pm100~\rm{M}_{\odot}$yr$^{-1}$. This is consistent with the processes that trigger quasars evolving in a similar way to the processes that trigger star formation. There is a rise in mean star formation rate from $z=2.5$ to $z=2.1$, which could be connected to evolution in the mass of the quasar hosts, but could also be due to selection effects in the DR9 catalog. 

2 - Higher rates of star formation are seen in more UV-luminous quasars, consistent with higher star formation rates correlating with higher accretion rates. We obtain the following relation between star formation rate and bolometric accretion luminosity

\begin{equation*}
\dot{\rm{M}}_s = (11\pm135) + (504\pm114) exp\left[\frac{(2.4\pm1.5)\times10^{12}}{L_{b}}\right]^{-1} 
\end{equation*}

\noindent This is consistent with a `maximal' {\itshape typical} star formation rate of $\sim600$M$_{\odot}$ yr$^{-1}$, in quasar hosts, perhaps due to saturation of the starburst by supernova winds. At lower star formation rates the relation between $\dot{\rm{M}}_s$ and black hole accretion rate is consistent with a power law with index $0.4 \lesssim \alpha \lesssim 1.1$; a linear fit gives:

\begin{equation*}
\frac{\dot{\rm{M}}_{bh}}{\dot{\rm{M}}_{s}} = \frac{1.91\pm0.27}{\eta}\times10^{-3} 
\end{equation*}

\noindent The existence of such a correlation in our study, when some previous studies have found no correlation, can be explained via a combination of two factors. First, that an $\dot{\rm{M}}_s - \dot{\rm{M}}_{bh}$ becomes stronger with increasing redshift, especially around $z=2$, corresponding to an epoch with a higher free gas fraction. Second, that short-term ($\lesssim100\,$Myr) AGN variability introduces scatter in the $\dot{\rm{M}}_s - L_{b}$ relation derived from measurements of objects individually, but which is averaged out in our stacking analyses. It is also plausible that very high or very low star formation rates and black hole accretion rates may not correlate with each other. This implies that a correlation between $\dot{\rm{M}}_s$ and $\dot{\rm{M}}_{bh}$ exists in certain parts of the $z - \dot{\rm{M}}_s - \dot{\rm{M}}_{bh}$ parameter space, dependent on the availability of free baryons, the trigger for activity, and the potential positive and negative effects of AGN on star formation.

3 - Higher rates of star formation are seen in quasars with more massive black holes. Assuming a relation of the form $\dot{\rm{M}}_s \propto \rm{M}_{bh}^{\alpha}$ and allowing slope and intercept to vary then an index of $0.2 \lesssim \alpha \lesssim 1.1$ fits the data. Further assuming that $\rm{M}_{bh} \propto FWHM_{\rm{CIV}}^{2} L_{UV}^{0.5}$, and with the dependencies on $z - \dot{\rm{M}}_s - \dot{\rm{M}}_{bh}$  parameter space noted above, then our results are consistent with star formation rates in quasars scaling with black hole mass. An additional, separate scaling with accretion rate is possible, but our results do not require it. A plausible physical origin is that star formation rates and black hole mass are both driven by the available gas reservoir.

4 - We see no clear relationship between star formation rate and Eddington ratio. There is a possibility that higher star formation rates are seen in systems with lower Eddington ratios, but the data are consistent with a flat relation. A flat relation is straightforward to understand if the underlying driver is the relation between star formation rate and black hole mass, since there would no reason to expect a relation between star formation rate and how {\itshape efficiently} the black hole is accreting. If however low $\lambda_{edd}$ values do correspond to higher star formation rates, then this may be due to smaller mass black holes accreting more efficiently, and/or that the peak star formation rate occurs some time before the peak in the AGN luminosity. 

5 - We see no clear relation between star formation rate and the asymmetry of the \ion{C}{iv} line. This relation is consistent with the idea that line asymmetries arise due to the relative orientation of the quasar; assuming that the infrared emission from star formation is optically thin then variations in line asymmetry would be accompanied by no net variation in star formation rate. There are two deviations from this flat relation, one consistent with AGN winds quenching star formation and one consistent with triggering, but both deviations are barely significant. 

6 - There is a decline in star formation rate with rising \ion{C}{iv} EW. The rise in star formation rate with rest-frame UV-luminosity suggests that part of this decline is a symptom of the Baldwin effect, but the dynamic range in star formation rate with \ion{C}{iv} EW is wider than the dynamic range in star formation rate with $M_{i}$. The most plausible explanation for this additional dynamic range is a contribution from three factors. First is that $M_i$ is not a linear tracer of L$_{2500}$. Second is that the Baldwin effect changes form at high $L_b$. Third is that high \ion{C}{iv} EW values signpost a change in the scaling relation between $\dot{\rm{M}}_s$ and $\dot{\rm{M}}_{bh}$.

\section{Acknowledgements}
We thank the referee for a very helpful report. {\itshape Herschel} is an ESA space observatory with instruments provided by European-led Principal Investigator consortia and with participation from NASA. The Herschel spacecraft was designed, built, tested, and launched under a contract to ESA managed by the Herschel/Planck Project team by an industrial consortium under the overall responsibility of the prime contractor Thales Alenia Space (Cannes), and including Astrium (Friedrichshafen) responsible for the payload module and for system testing, Thales Alenia Space (Turin) responsible for the service module, and Astrium (Toulouse) responsible for the telescope, with in excess of a hundred subcontractors. SPIRE has been developed by a consortium of institutes led by Cardiff Univ. (UK) and including: Univ. Lethbridge (Canada); NAOC (China); CEA, LAM (France); IFSI, Univ. Padua (Italy); IAC (Spain); Stockholm Observatory (Sweden); Imperial College London, RAL, UCL-MSSL, UKATC, Univ. Sussex (UK); and Caltech, JPL, NHSC, Univ. Colorado (USA). This development has been supported by national funding agencies: CSA (Canada); NAOC (China); CEA, CNES, CNRS (France); ASI (Italy); MCINN (Spain); SNSB (Sweden); STFC, UKSA (UK); and NASA (USA). Funding for SDSS-III has been provided by the Alfred P. Sloan Foundation, the Participating Institutions, the National Science Foundation, and the U.S. Department of Energy Office of Science. SDSS-III is managed by the Astrophysical Research Consortium for the Participating Institutions of the SDSS-III Collaboration including the University of Arizona, the Brazilian Participation Group, Brookhaven National Laboratory, Carnegie Mellon University, University of Florida, the French Participation Group, the German Participation Group, Harvard University, the Instituto de Astrofisica de Canarias, the Michigan State/Notre Dame/JINA Participation Group, Johns Hopkins University, Lawrence Berkeley National Laboratory, Max Planck Institute for Astrophysics, Max Planck Institute for Extraterrestrial Physics, New Mexico State University, New York University, Ohio State University, Pennsylvania State University, University of Portsmouth, Princeton University, the Spanish Participation Group, University of Tokyo, University of Utah, Vanderbilt University, University of Virginia, University of Washington, and Yale University. AF acknowledges support from the ERC via an Advanced Grant 321323-NEOGAL.

\bsp	

\label{lastpage}
\end{document}